\newif{\ifjournal}
  \journaltrue

\ifjournal
  \documentclass[usenatbib]{mn2e}
  \usepackage{graphicx,mathptm}
  \newcommand{\gtrsim}{\ga} 
  \newcommand{\lesssim}{\la} 
\else
  \documentclass{paper}
  \newcommand{\ga}{\gtrsim} 
  \newcommand{\la}{\lesssim} 
\fi

\renewcommand{\d}{\mathrm{d}}

\begin{document}  
  
\title{The impact of cluster mergers on arc statistics}
 
\ifjournal\author[Torri et al.]
  {Elena Torri$^{1}$, Massimo Meneghetti$^{1}$, Matthias
   Bartelmann$^{2,3}$, \newauthor Lauro Moscardini$^4$, Elena
   Rasia$^1$, Giuseppe Tormen$^{1}$ \\
   $^1$Dipartimento di Astronomia, Universit\`a di Padova, vicolo
   dell'Osservatorio 2, I--35122 Padova, Italy\\
   $^2$Max-Planck-Institut f\"ur Astrophysik,
   Karl-Schwarzschild-Strasse 1, D--85748 Garching, Germany\\
   $^3$Institut f\"ur Theoretische Astrophysik, Tiergartenstr.~15,
   D--69121 Heidelberg, Germany\\
   $^4$Dipartimento di Astronomia, Universit\`a di Bologna, via
   Ranzani 1, I--40127 Bologna, Italy}
\else\author
  {Elena Torri$^{1}$, Massimo Meneghetti$^{1}$, Matthias
   Bartelmann$^{2,3}$, \newauthor Lauro Moscardini$^4$, Elena
   Rasia$^1$, Giuseppe Tormen$^{1}$ \\
   $^1$Dipartimento di Astronomia, Universit\`a di Padova, vicolo
   dell'Osservatorio 2, I--35122 Padova, Italy\\
   $^2$Max-Planck-Institut f\"ur Astrophysik,
   Karl-Schwarzschild-Strasse 1, D--85748 Garching, Germany\\
   $^3$Institut f\"ur Theoretische Astrophysik, Tiergartenstr.~15,
   D--69121 Heidelberg\\
   $^4$Dipartimento di Astronomia, Universit\`a di Bologna, via
   Ranzani 1, I--40127 Bologna, Italy}
\fi

\date{Accepted 2003 ???? ???; Received 2003 ???? ???;  
in original form 2003 ???? ??}  
  
\ifjournal\maketitle\fi
  
\begin{abstract}
We study the impact of merger events on the strong lensing properties
of galaxy clusters. Previous lensing simulations were not able to
resolve dynamical time scales of cluster lenses, which arise on time
scales which are of order a Gyr. In this case study, we first describe
qualitatively with an analytic model how some of the lensing
properties of clusters are expected to change during merging
events. We then analyse a numerically simulated lens model for the
variation in its efficiency for producing both tangential and radial
arcs while a massive substructure falls onto the main cluster body. We
find that: (1) during the merger, the shape of the critical lines and
caustics changes substantially; (2) the lensing cross sections for
long and thin arcs can grow by one order of magnitude and reach their
maxima when the extent of the critical curves is largest; (3) the
cross section for radial arcs also grows, but the cluster can
efficiently produce this kind of arcs only while the merging
substructure crosses the main cluster centre; (4) while the arc cross
sections pass through their maxima as the merger proceeds, the
cluster's X-ray emission increases by a factor of $\sim5$. Thus, we
conclude that accounting for these dynamical processes is very
important for arc statistics studies. In particular, they may provide
a possible explanation for the arc statistics problem.
\end{abstract}

\ifjournal\else\maketitle\fi

\section{Introduction}

The abundance of strong gravitational lensing events produced by
galaxy clusters is determined by several factors. Since light
deflection depends on the distances between observer, lens and source,
gravitational lensing effects depend on the geometrical properties of
the universe. On the other hand, gravitational arcs are rare events
caused by a highly nonlinear effect in cluster cores. They are thus
not only sensitive to the number density of galaxy clusters, but also
to their individual internal structure. Since these factors depend on
cosmology and in particular on the present value of the matter density
parameter $\Omega_\mathrm{0M}$ and the contribution from the
cosmological constant $\Omega_{0\Lambda}$, arc statistics establishes
a highly sensitive link between cosmology and our understanding of
cluster formation.

Using the ray-tracing technique for studying the lensing properties of
galaxy clusters obtained from N-body simulations, \citet{BA98.2}
showed that the number of \emph{giant arcs}, commonly defined as arcs
with length-to-width ratio larger than $10$ and $B$-magnitude smaller
than $21.5$ \citep{WU93.1}, which is expected to be detectable on the
whole sky, differs by orders-of-magnitudes between high- and
low-density universes, strongly depending even on the cosmological
constant. In particular, they estimated that the number of such arcs
in a $\Lambda$CDM cosmological model ($\Omega_\mathrm{0M}=0.3$,
$\Omega_{0\Lambda}=0.7$) is of the order of some hundreds on the whole
sky, while roughly ten times more arcs are expected in an OCDM
cosmological model ($\Omega_\mathrm{0M}=0.3$, $\Omega_{0\Lambda}=0$).

Although still based on limited samples of galaxy clusters,
observations indicate that the occurrence of strong lensing events on
the sky is rather high \citep{LU99.1,ZA03.1,GL03.1}. For example,
searching for giant arcs in a sample of X-ray luminous clusters
selected from the \emph{Einstein Observatory} Extended Medium Survey,
\citet{LU99.1} found that their frequency is about $0.2-0.4$ arcs per
massive cluster. Despite the obvious uncertainties in the
observations, the only cosmological model for which the number of
giant arcs expected from numerical simulations of gravitational
lensing comes near the observed number is the OCDM model. In
particular the observed incidence of strongly lensed galaxies exceeds
the predictions of a $\Lambda$CDM model by about a factor of ten. On
the other hand, based on the observations of type Ia supernovae
\citep{PE99.1,TO03.2} and the recent accurate measurements of the temperature
fluctuations of the Cosmic Microwave Background obtained with balloon
experiments (e.g.~\citealt{ST01.1,JA01.1,AB02.1,BE03.2}) or by the
$WMAP$ mission (e.g.~\citealt{BE03.1,SP03.1}), the $\Lambda$CDM model
has become the favourite cosmogony. This is known as the \emph{arc
statistics problem}: the mismatch between the observed number of arcs
and the number expected in the $\Lambda$CDM model preferred by
virtually all other cosmological experiments hints at a lack of
understanding of cluster formation.

Several extensions and improvements of the numerical simulations
failed in finding a solution to this problem in the lensing
simulations. \citet{ME00.1} studied the influence of individual
cluster galaxies on the ability of clusters to form large
gravitational arcs, finding that their effect is statistically
negligible. \citet{CO99.2} and \citet{KA00.1}, using spherical
analytic models and the Press-Schechter formalism for modelling the
lens cluster population, predicted a weaker dependence of arc
statistics on $\Omega_{0\Lambda}$, but \citet{ME03.1}, comparing
numerical models of galaxy clusters and their analytical fits, showed
that analytic models are inadequate for quantitative studies of arc
statistics. Finally, \citet{ME03.2} found that the presence of central
cD galaxies can increase the cluster efficiency for producing giant
arcs by not more than a factor of about two.

As alternative solution of the arc statistics problem, \citet{BA03.1}
recently investigated arc properties in cosmological models with more
general forms of dark energy than a cosmological constant. Several
studies showed that haloes should be more concentrated in these models
than in cosmological-constant models with the same dark energy density
today (\citealt{BA02.1,MA03.1,KL03.1,DO03.1}), allowing them to be
more efficient for strong lensing. Using simple models with constant
quintessence parameter, \citet{BA03.1} found that the relative change
of the halo concentration is not sufficient to produce an increment of
one order of magnitude in the expected number of giant
arcs. Nonetheless, other more elaborate dark-energy models need to be
evaluated numerically and will be discussed in a future paper
(Meneghetti et al.~in preparation).

In this paper we investigate another possible effect which could not
be properly considered in the previously mentioned numerical
simulations of gravitational lensing by galaxy clusters. In those
works, the lensing cross sections for giant arcs of each numerical
model were evaluated at different redshifts, with a typical time
separation between two consecutive simulation outputs of approximately
$\Delta t \sim 1$ Gyr. Therefore all the dynamical processes arising
in the lenses on time scales smaller than $\Delta t$ were not
resolved.

N-body simulations show that dark matter haloes of different masses
continuously fall onto rich clusters of galaxies \citep{TO97.1}. The
typical time scale for such events is $\sim1-2$ Gyr, which therefore
might be too short for having been properly accounted for in the
previous lensing simulations.

However, the effects of mergers on the lensing properties of galaxy
clusters may potentially be very important.  As discussed by
\citet{BA95.1} and \citet{ME03.1}, substructures play a very important
role for determining the cluster efficiency for lensing. Indeed,
analytic models, where substructures and asymmetries in the lensing
mass distributions are not properly taken into account, systematically
underestimate the lensing cross sections of the numerical models. The
main reason is that mass concentrations around and within clusters
enhance the shear field, increasing the length of the critical curves,
and consequently the probability of forming long arcs becomes higher.

Given the strong impact of substructures on the lensing properties of
galaxy clusters, it is reasonable to expect that during the passage of
a massive mass concentration through or near the cluster centre, the
lensing efficiency might sensitively fluctuate. While the substructure
is approaching the main cluster clump, the intensity of the shear
field and, consequently, the shape of the critical curves may
substantially change. Moreover, while the infalling dark matter halo
gets closer to the cluster centre, the projected surface density
increases, making the cluster much more efficient for strong lensing.

This paper describes a case study on how much the lensing cross
sections change during the infall of a massive dark matter halo on the
main cluster progenitor. For doing so, we investigate the lensing
properties of a numerically simulated galaxy cluster during a redshift
interval when a major merging event occurs. The general aim is to
understand if mergers can enhance the cluster lensing efficiency
sufficiently to provide a solution to the arc statistics problem.

The plan of the paper is as follows. In Section 2 we use a simple
analytic model based on the NFW density profile for computing the
strong lensing properties produced by cluster mergers.  In Section 3
we describe our lensing simulations utilizing a numerical model
obtained from a high-resolution N-body simulation; a method for
evaluating the dynamical state of the cluster is also introduced.  In
Section 4 we present our results for critical lines and caustics, and
we estimate the expected number of tangential and radial arcs.
Section 5 is devoted to a discussion of the observational implications
of our results, and we summarize the results and present our
conclusions in Sect.~6.

\section{Analytic models}

We begin by investigating with the help of analytic models how the
lensing properties of a cluster lens are expected to change while a
massive substructure passes through its centre. We will model both the
main cluster clump and the infalling substructure as spherical bodies
with the density profile found in numerical simulations by \citet[
hereafter NFW]{NA96.1}.

\subsection{NFW model}

The NFW density profile is given by 
\begin{equation}
  \rho(r)=\frac{\rho_\mathrm{s}}
  {(r/r_\mathrm{s})(1+r/r_\mathrm{s})^2}\;,
  \label{eq:nfw}
\end{equation}
where $\rho_\mathrm{s}$ and $r_\mathrm{s}$ are the characteristic
density and the distance scale, respectively (see
\citealt{NA97.1}). The logarithmic slope of this density profile
changes from $-1$ at the centre to $-3$ at large radii.

We briefly summarize here the main features of NFW haloes. First,
$\rho_\mathrm{s}$ and $r_\mathrm{s}$ are not independent. Second, the
ratio between the radius $r_{200}$, within which the mean halo density
is $200$ times the critical density, and $r_s$ defines the halo
concentration, $c\equiv r_{200}/r_\mathrm{s}$. Numerical simulations
show that this parameter systematically changes with the halo virial
mass, which is thus the only free parameter. Several algorithms exist
for computing the concentration parameter from the halo virial mass
\citep{NA97.1,EK01.1,BU01.1}. In the following analysis we adopt the
method proposed by \citet{NA97.1}. Third, the concentration also
depends on the cosmological model, implying that the lensing
properties of haloes with identical mass are different in different
cosmological models if they are modelled as NFW spheres. The lensing
properties of haloes with NFW profiles have been discussed in several
papers
\citep{BA96.1,WR00.1,WY01.1,LI02.1,PE02.1,BA02.1,ME03.2,ME03.1}.

Due to axial symmetry, lensing by an NFW sphere reduces to a
one-dimensional problem. We define the optical axis as the straight
line passing through the observer and the centre of the lens, and
introduce the physical distances perpendicular to the optical axis on
the lens and source planes, $\xi$ and $\eta$, respectively. We
introduce dimension-less coordinates $x=\xi/\xi_0$ and $y=\eta/\eta_0$
by means of the coordinate scales $\xi_0=r_\mathrm{s}$ and
$\eta_0=(D_\mathrm{s}/D_\mathrm{l})\,\eta_0$, where $D_\mathrm{s}$ and
$D_\mathrm{l}$ are the angular diameter distances to the source and
lens planes, respectively.

\begin{figure*}
\centering
  \includegraphics[width=.25\hsize]{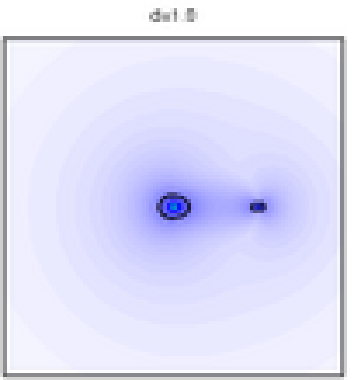}
  \includegraphics[width=.25\hsize]{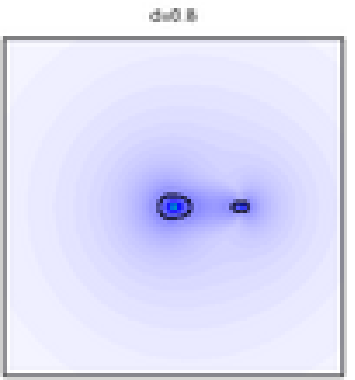}
  \includegraphics[width=.25\hsize]{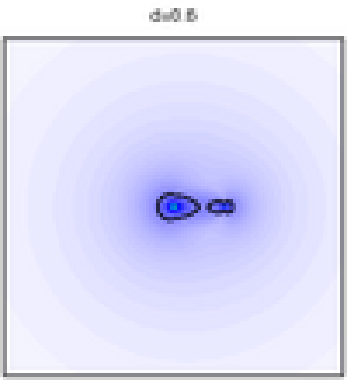}
  \includegraphics[width=.25\hsize]{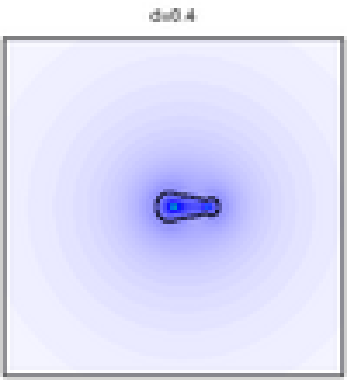}
  \includegraphics[width=.25\hsize]{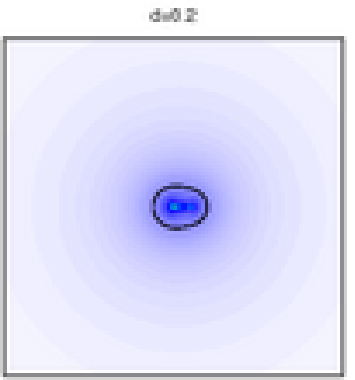}
  \includegraphics[width=.25\hsize]{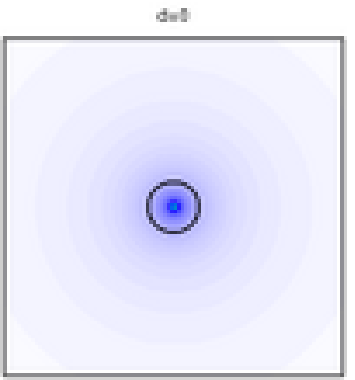}
\caption{Illustration of the analytic model described in
  Sect.~\ref{section:mergexpectations}: Maps of the tangential
  eigenvalue $\lambda_t$ for all configurations of the main cluster
  body relative to the infalling substructure. The critical lines are
  over-plotted in each map. The (comoving) size of each panel is
  $\sim1.2\,h^{-1}\,\mathrm{Mpc}$. Different panels refer to different
  distances $d$ (in units of the scale radius of the larger halo)
  between the two haloes, as indicated by the top label.}
\label{figure:critl}
\end{figure*}

Using this notation, the density profile (\ref{eq:nfw}) implies the
surface mass density \citep{BA96.1}
\begin{equation}
  \Sigma(x)=\frac{2 \rho_s r_s}{x^2-1}f(x) \ ,
\end{equation}
with 
\begin{equation}
f(x)= \left\{
\begin{array}{r@{\quad \quad}l}
1-\frac{2}{\sqrt{x^2-1}} \ \mbox{arctan} \sqrt{\frac{x-1}{x+1}} & (x>1) \\
1-\frac{2}{\sqrt{1-x^2}} \ \mbox{arctanh}
\sqrt{\frac{1-x}{1+x}} & (x<1) \\ 0 & (x=1)
\end{array} \right.
\ . 
\end{equation}
The lensing potential is given by
\begin{equation}
  \Psi(x)=4\kappa_\mathrm{s} g(x)\:,
\label{eq:nfwPsi}
\end{equation}
where
\begin{equation}
g(x) = \frac{1}{2}\ln^2 \frac{x}{2} + 
 \left\{
\begin{array}{r@{\quad \quad}l}
  2\,\mbox{arctan}^2\sqrt{\frac{x-1}{x+1}} & (x>1) \\
  -2\,\mbox{arctanh}^2\sqrt{\frac{1-x}{1+x}} & (x<1) \\
0 & (x=1)
\end{array} \right.
\;,
\end{equation}  
and $\kappa_s \equiv \rho_s r_s \Sigma_\mathrm{cr}^{-1}$. here
$\Sigma_\mathrm{cr}=[c^2/(4\pi
G)]\,[D_\mathrm{s}/(D_\mathrm{l}D_\mathrm{ls})]$ is the critical
surface mass density for strong lensing and $D_\mathrm{ls}$ is the
angular diameter distance between lens and source. This implies the
deflection angle
\begin{equation}
  \alpha(x)=\frac{\d\psi}{\d x}=\frac{4\kappa_\mathrm{s}}{x} h(x) \:,
  \label{eq:nfwdefang}
\end{equation}
with
\begin{equation}
h(x) = \ln \frac{x}{2} +
 \left\{
\begin{array}{r@{\quad \quad}l}
  \frac{2}{\sqrt{x^2-1}}\,
  \mbox{arctan}\sqrt{\frac{x-1}{x+1}} & (x>1) \\
  \frac{2}{\sqrt{1-x^2}}\,
  \mbox{arctanh}\sqrt{\frac{1-x}{1+x}} & (x<1) \\
  1 & (x=1)
\end{array} \right.
\;.
\end{equation}

\subsection{Expectations}
\label{section:mergexpectations}

\begin{figure*}
\centering
  \includegraphics[width=.35\hsize]{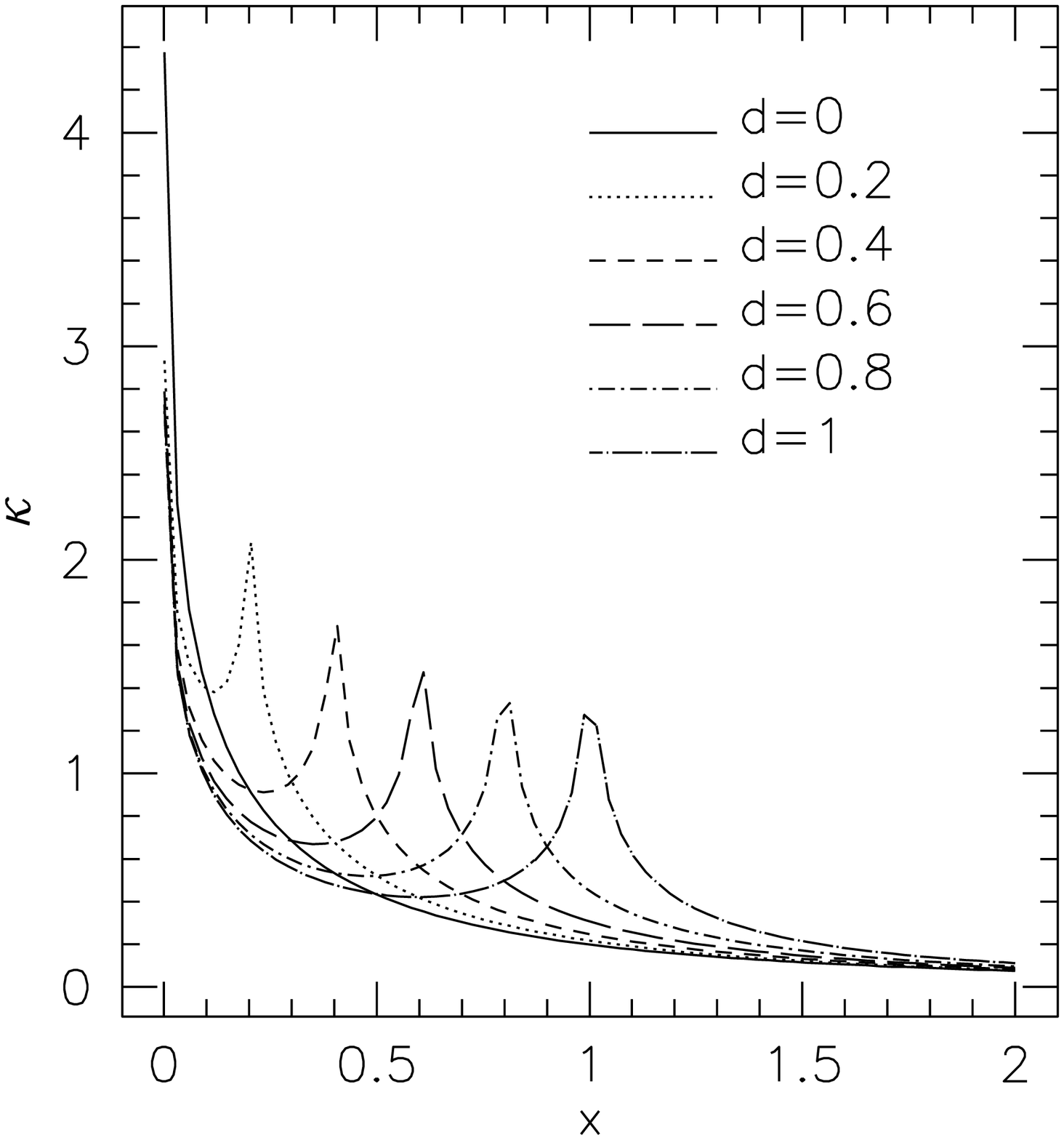}
  \includegraphics[width=.35\hsize]{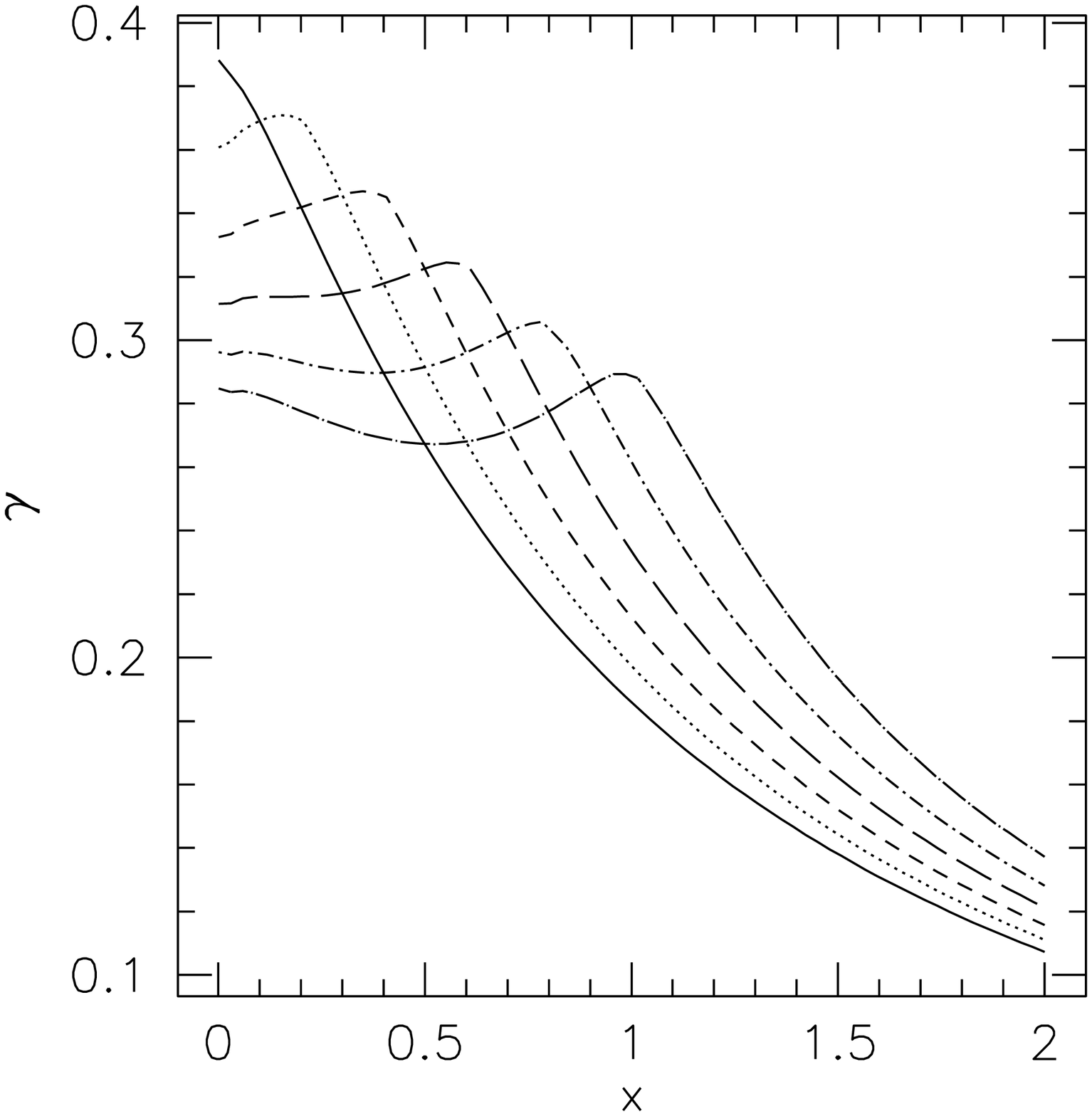}
\caption{Profiles of convergence $\kappa$ (left panel) and shear
  $\gamma$ (right panel) along the axis connecting the centres of the
  main cluster body and the infalling substructure illustrated in
  Fig.~\ref{figure:critl}. Different curves refer to different
  distances $d$ (in units of the scale radius of the larger halo)
  between the two haloes. Notice the increased shear level at fixed
  $x$ during the merger.}
\label{figure:kgplots}
\end{figure*}

\begin{figure*}
\centering
  \includegraphics[width=0.8\hsize]{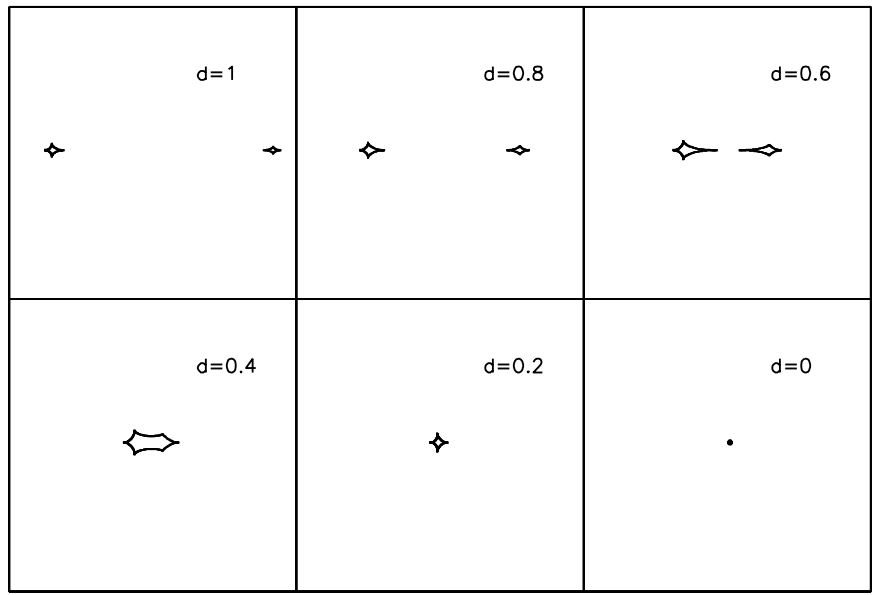}
\caption{Caustics for all configurations of the main cluster body
  relative to the infalling substructure shown in
  Fig.~\ref{figure:critl}. The size of each panel is $\sim 200 \
  h^{-1}$kpc on a side.  Different panels refer to different distances
  $d$ (in units of the scale radius of the larger halo) between the
  two haloes.}
\label{figure:caustl}
\end{figure*}

\begin{figure*}
\centering
  \includegraphics[width=.35\hsize]{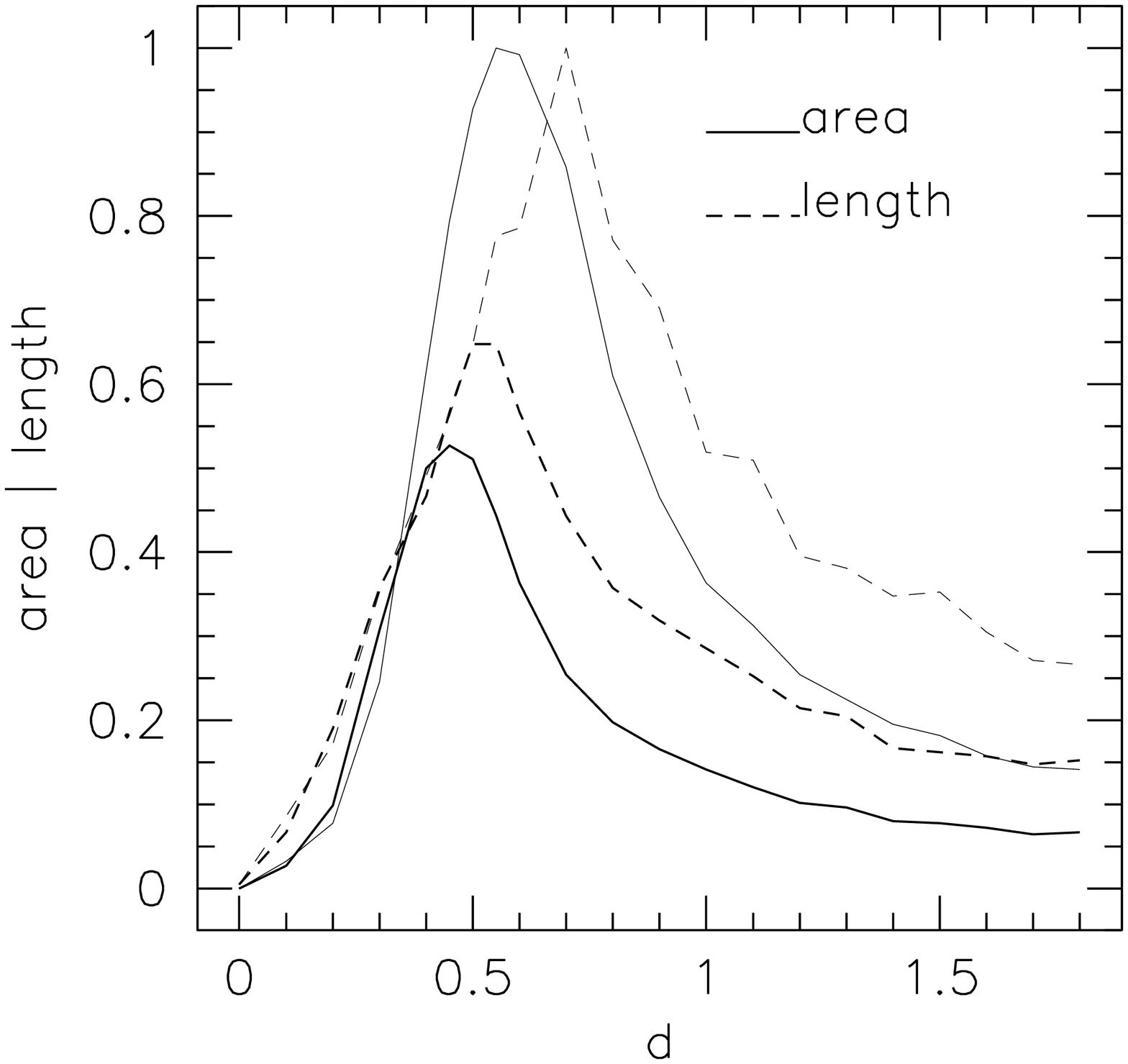}
  \includegraphics[width=.35\hsize]{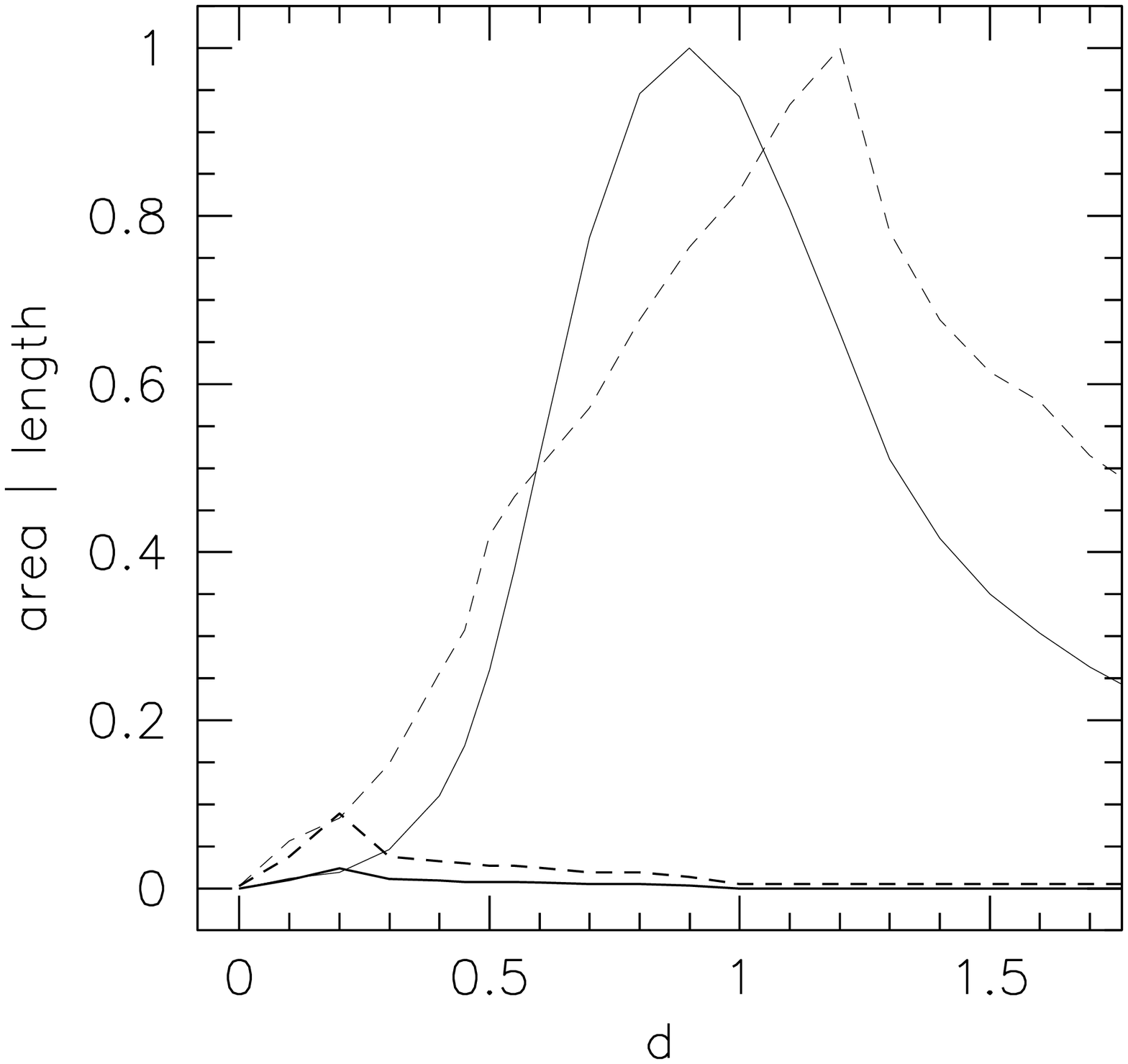}
\caption{Left panel: caustic area and length (solid and dashed lines,
  respectively) as functions of the distance between the merging haloes
  (in units of the scale radius of the more massive halo). The lens
  redshift is $z_{\rm l}=0.3$. Thick and thin lines refer to source
  redshifts $z_{\rm s}=1$ and $z_{\rm s}=2$, respectively. Curves are
  normalized to the maximum value of area and length for sources at
  $z_{\rm s}=2$. Right panel: as in the left panel, but for lenses at
  redshift $z_{\rm l}=0.8$}
\label{figure:arealength}
\end{figure*}

In this analytic computation, the main cluster body has a mass of
$M_\mathrm{main}=10^{15} \ h^{-1} M_{\odot}$. It merges with a massive
substructure of mass $M_\mathrm{sub}=M_\mathrm{main}/4=2.5 \times
10^{14} h^{-1} M_{\odot}$, which is a fairly typical ratio for merging
halo masses. In this simple test, we keep the lens plane at the fixed
redshift $z_\mathrm{l}=0.3$, while the source plane is placed at
redshift $z_\mathrm{s}=1$. We assume a background $\Lambda$CDM
cosmological model, with $\Omega_\mathrm{0M}=0.3$,
$\Omega_{0\Lambda}=0.7$, and Hubble constant (in units of 100
km/s/Mpc) $h=0.7$.

We simulate the infall of the substructure onto the main clump,
starting from the initial configuration when the smaller halo is
placed at distance $r_s$ from the larger one, where $r_s$ is the scale
radius of the most massive clump. For a halo of mass $10^{15} h^{-1}
M_{\odot}$ in a $\Lambda$CDM model, $r_\mathrm{s}$ corresponds to
$\sim 310 \ h^{-1}$kpc. Then, we reduce the distance between the two
haloes by moving the smaller towards the larger one by $0.2r_s$ per
time step, until the two haloes are exactly aligned.

For any new configuration, we trace a bundle of $512 \times 512$ light
rays through a regular grid on the lens plane which covers a region
whose side length is $4r_s$. This is large enough to encompass the
critical lines of both the clumps.

Using (\ref{eq:nfwdefang}), we compute the contributions $\vec
\alpha^\mathrm{main}$ and $\vec \alpha^\mathrm{sub}$ of both the main
clump and the substructure to the deflection angle of each ray. Being
linear in mass, the total deflection angle of a ray passing through
the mass distribution is the sum of the contributions from each mass
element of the deflector. Thus, the deflection angle of each ray
parametrized by its grid indices $(i,j)$ on the lens plane is given by
\begin{equation} 
  \vec\alpha_{ij} = \vec\alpha^\mathrm{main}_{ij} +
  \vec\alpha^\mathrm{sub}_{ij} \ .
\end{equation} 

The convergence and the shear maps of the lens system are
reconstructed from the deflection angles. The true position of the
source on the source plane, $\vec{y}$, and its observed position on
the lens plane, $x$, are related by the lens equation
\begin{equation}
\vec{y}=\vec{x}-\vec{\alpha}(\vec{x}) \ .
\label{eq:lens}
\end{equation}
The local properties of the lens mapping are described by the Jacobian
matrix of (\ref{eq:lens}),
\begin{equation}
  A_{hk} \equiv \frac{\partial y_h}{\partial x_k} =
  \delta_{hk}-\frac{\partial \alpha_h}{\partial x_k} \ ,
\end{equation}
where $\delta_{hk}$ is the Kronecker symbol. The convergence $\kappa$
and the shear components $\gamma_1$ and $\gamma_2$ are found from
$A_{hk}$ through the standard relations
\begin{eqnarray}
  \kappa(\vec x) & = & \ \ \  \frac{1}{2} [A_{11}(\vec x)+A_{22}(\vec
  x)] \ , \\ 
  \gamma_1(\vec x) & = & -\frac{1}{2} [A_{11}(\vec x)-A_{22}(\vec x)]
  \ , \\
  \gamma_2(\vec x) & = & -\frac{1}{2} [A_{12}(\vec x)+A_{21}(\vec x)] \ .
\end{eqnarray}
The Jacobian matrix is symmetric and can be diagonalised. Its two
eigenvalues are
\begin{eqnarray}
  \lambda_\mathrm{t} = 1-\kappa-\gamma & \mbox{and} & \lambda_\mathrm{r}=
  1 -\kappa +\gamma \ ,
\end{eqnarray}
where $\gamma=\sqrt{\gamma_1^2+\gamma_2^2}$.  Radial and tangential
critical lines are located where the conditions
\begin{eqnarray}
  \lambda_\mathrm{t}=0 & \mbox{and} & \lambda_\mathrm{r}=0 
\end{eqnarray}
are satisfied, respectively. The corresponding caustics in the source
plane, close to which sources are imaged as large arcs, are obtained
by applying the lens equation to the critical curves.

We are particularly interested in understanding by how much the
cluster efficiency for producing tangential arcs changes during the
merger phase. The size of the lensing cross
sections for the formation of tangential arcs is strictly connected to
the extent of the tangential critical curves and caustics: the longer
the critical curves and the corresponding caustics are, the larger are
the lensing cross sections. Therefore, we first investigate the shape of the
tangential critical lines and caustics. In Fig.~\ref{figure:critl}, we
show the maps of the tangential eigenvalue $\lambda_\mathrm{t}$ for
all configurations of the main clump relative to the infalling
substructure. In each map, we over-plot the tangential critical
lines. As the two haloes approach each other, their critical lines
widen and stretch along the approaching direction. When the distance
is $\sim 0.6r_s$, the critical lines touch and merge to form a single
critical line. While the substructure overlaps with the main cluster
body, it shrinks in the direction along which the merger proceeds,
while it widens perpendicular to it.

\begin{figure*}
\flushleft
  \includegraphics[width=0.24\hsize]{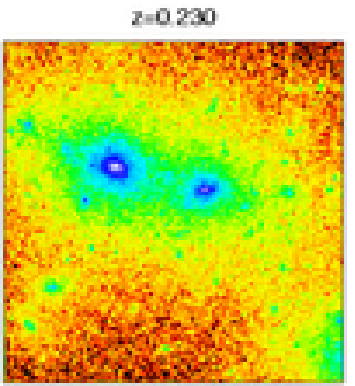}
  \includegraphics[width=0.24\hsize]{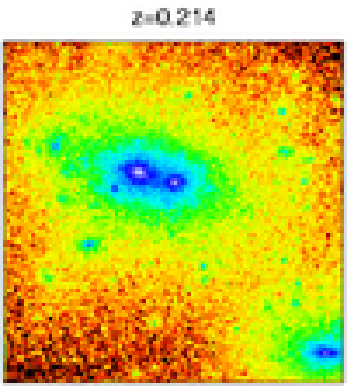}
  \includegraphics[width=0.24\hsize]{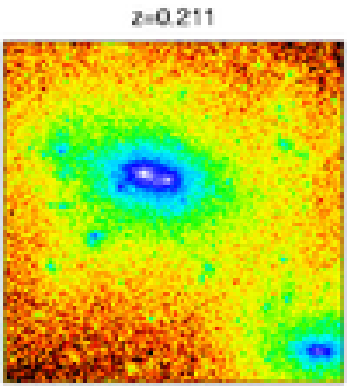}
  \includegraphics[width=0.24\hsize]{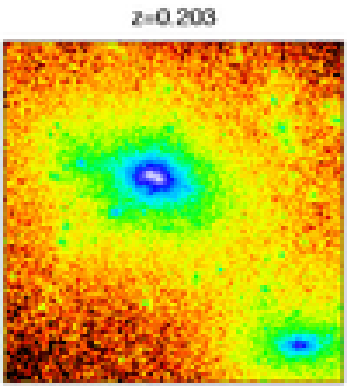}
  \includegraphics[width=0.24\hsize]{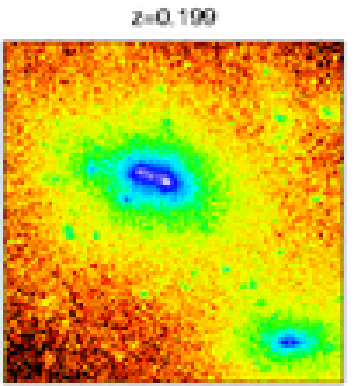}
  \includegraphics[width=0.24\hsize]{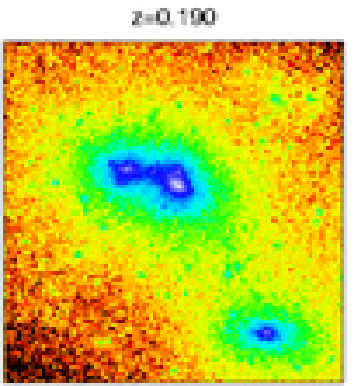}
  \includegraphics[width=0.24\hsize]{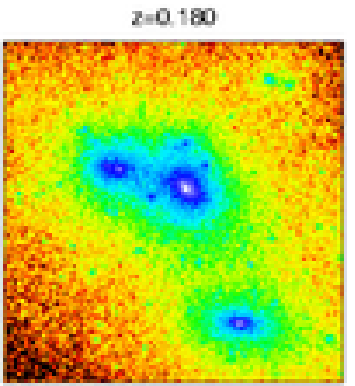}
\caption{Surface density maps of the numerically simulated cluster at
  several redshifts between $z=0.230$ and $z=0.180$, during the
  merger. The scale of each panel is $3 \ h^{-1}$Mpc.}
\label{figure:mergsurfmaps}
\end{figure*}

We can explain the evolution of the critical lines as follows. As the
two haloes approach each other, the shear in the region between them
grows. This stretches the critical lines in the direction along which
the two haloes are approaching. When the critical lines merge, the
shear continues growing in the region inside the unique critical
line. At this point the critical curve stops growing in a privileged
direction. It shrinks only in the direction along which the haloes
merge because of the decreasing distance between their centres. On the
other hand, it expands almost isotropically by the effect of the
increasing convergence. We show in Fig.~\ref{figure:kgplots} how the
convergence and the shear evolve along the axis connecting the halo
centres, where shrinking and stretching effects are most evident. The
main clump is kept at $x=0$. Different curves show the results for
different distances $d$ between the two haloes, in units of the scale
radius of the largest clump.

The caustics are shown in Fig.~\ref{figure:caustl}. Their evolution
reflects that of the critical lines. As the shear grows between the
two clumps, they are stretched and develop cusps. Then, after they
merge, the single caustic shrinks and reduces to a point when the two
haloes are exactly aligned. 

In the case of merging haloes, where very complicated structures
develop in the caustics, it is difficult to say whether the cross
sections for long and thin arcs scale as the area enclosed by the
caustics or as the caustic length. In Fig.~\ref{figure:arealength} we
show the evolution of both the area (solid lines) and the length
(dashed lines) of the caustics while the merger proceeds. Thick lines
in the left panel refer to the configuration of lenses and sources
described earlier. For completeness, in the same panel we also show 
the corresponding curves for sources at redshift $z_{\rm s}=2$ (thin
lines). In the right panel, the corresponding curves for lenses at
$z_{\rm l}=0.8$ are shown. As expected, the size of the caustics is
larger for sources more distant from the lens because of its increased
convergence. All curves in each panel are normalized to the maximum
value of length and enclosed area for sources at redshift $z_{\rm
s}=2$. Of course the area enclosed by the caustics grows more steeply
compared to the length of the caustics. It is interesting that
caustics reach their maximum length before they reach their maximum
area. This is due to the stretching of the caustics of the two
individual merging haloes towards each other: the cusps become very
peaked and pronounced until they touch, with large change of
length. The maximum value of the area enclosed is reached only shortly
thereafter, when the caustics have already started to enlarge
isotropically.

\section{Numerical methods}

\subsection{Numerical model}
\label{section:mergnummod}

We now turn to a numerical cluster model in order to quantify how the
strong lensing efficiency changes during a merger of a realistic
cluster mass distributions embedded into the tidal field of the
surrounding large-scale structure.

The numerical model we investigate here is part of a set of 17 objects
obtained using the technique of re-simulating at higher resolution a
patch of a pre-existing cosmological simulation. The re-simulation
method is described in \citep{TO97.1}. A detailed discussion of the
dynamical properties of the whole sample of these simulated clusters
is presented elsewhere \citep{TO03.1,RA03.1}.

The parent N-body simulation, with $512^3$ particles in a box of side
$479h^{-1}$ Mpc, has been produced by N.~Yoshida for the Virgo
Consortium (see also \citealt{JE98.1}).  It assumes a flat universe
with $\Omega_\mathrm{0M} = 0.3$, $\Omega_{0\Lambda} = 0.7$ and $h =
0.7$. The initial conditions correspond to a cold dark matter power
spectrum normalised to have $\sigma_8 = 0.9$ today. The particle mass
is $6.86 \times 10^{10}h^{-1} M_\odot$, which allows to resolve a
cluster-sized halo by several thousand particles; the gravitational
softening is $15h^{-1}$ kpc.  From the final output of this simulation
we randomly extracted some spherical regions containing a
cluster-sized dark matter halo.  Each of these regions was re-sampled
to build new initial conditions for a higher number of particles - on
average $10^6$ dark matter particles and the same number of gas
particles.  The mass of the dark matter particles is $\sim 5 \times
10^9 \ h^{-1} M_{\odot}$, while the mass of the gas particles is $\sim
10$ times smaller. The softening length for dark matter and gas is
$3.6 \ h^{-1}$kpc and $7.1 \ h^{-1}$kpc, respectively.  These initial
conditions were evolved using the publicly available code GADGET
\citep{SP01.1} from a starting redshift $z_\mathrm{in} = 35-50$ to
redshift zero.

In our sample of 17 objects we selected a simulated cluster undergoing
a major merger at $z\approx 0.2$. At redshift $z \sim 0.25$, when
their viral regions merge, the main cluster clump and the infalling
substructure have virial masses of $\sim 7 \times 10^{14} h^{-1}
M_{\odot}$ and $\sim 3 \times 10^{14} h^{-1} M_{\odot}$,
respectively. In order to have a very good time resolution for
resolving the complete merger in detail, we decided to re-simulate the
cluster between $z=0.25$ and $z=0.15$, obtaining 101 equidistant
outputs which we use for our following lensing analysis.

\begin{figure*}
\flushleft
  \includegraphics[width=0.33\hsize]{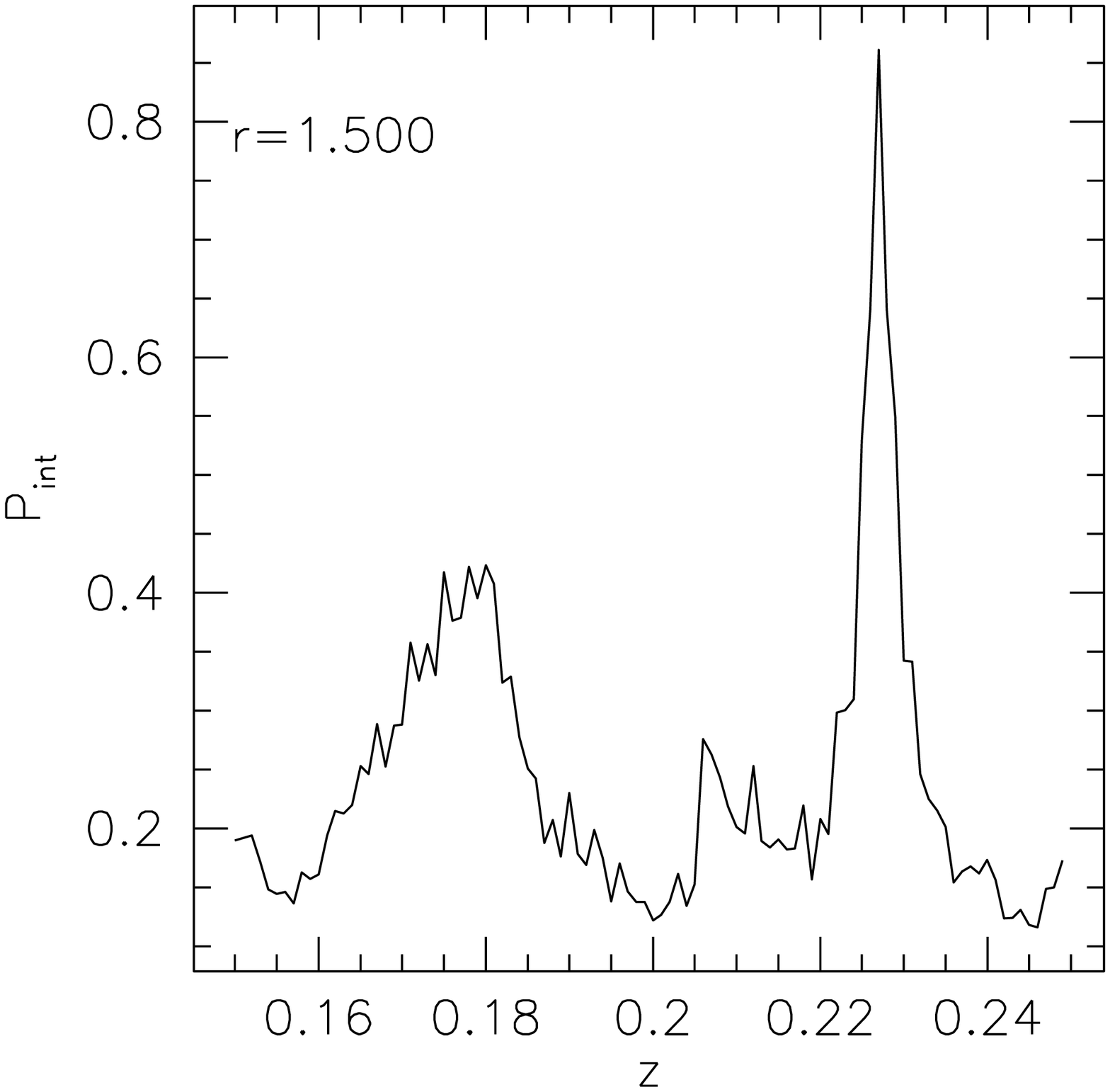}
  \includegraphics[width=0.33\hsize]{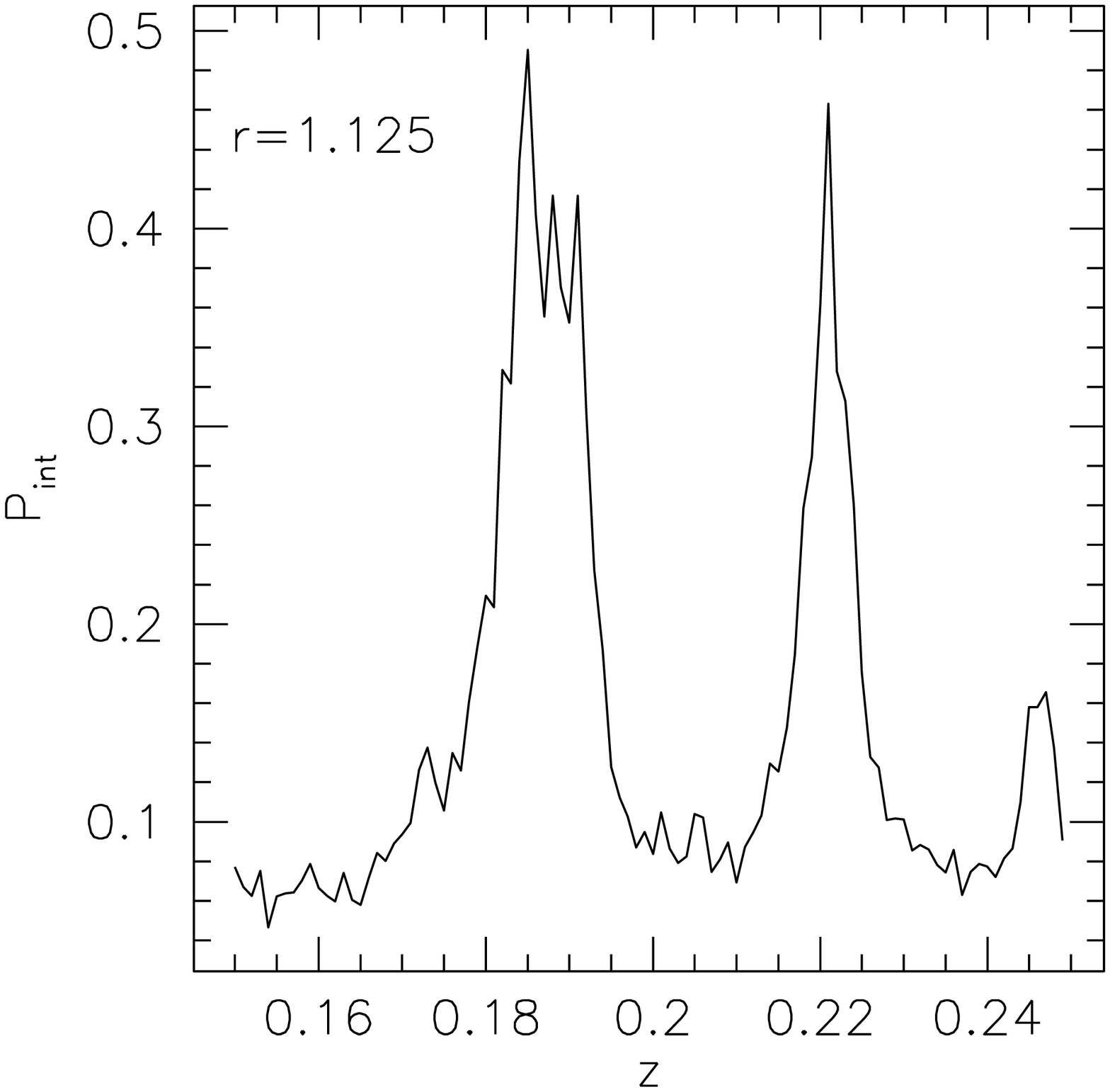}
  \includegraphics[width=0.33\hsize]{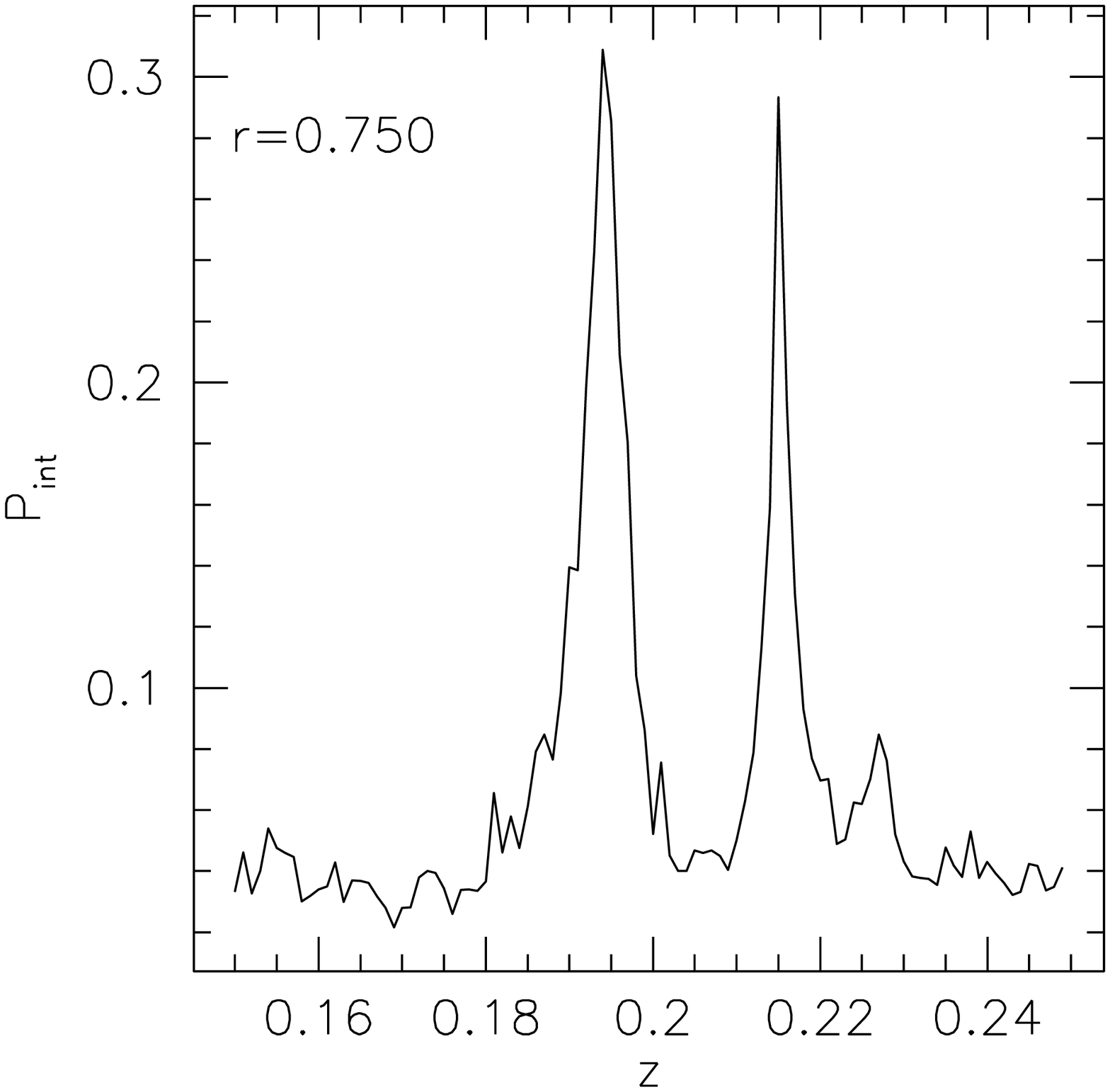}
  \includegraphics[width=0.33\hsize]{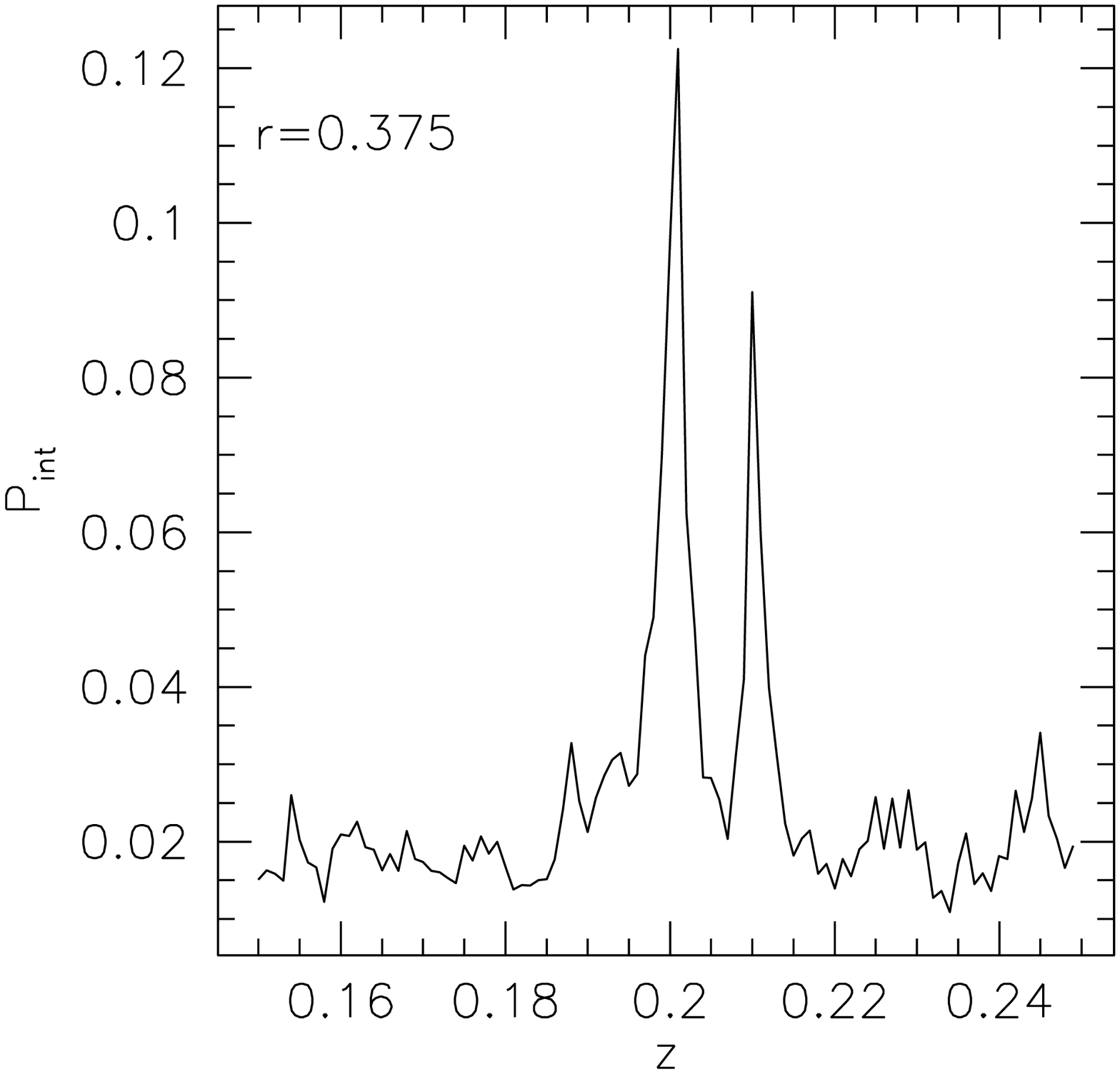}
  \includegraphics[width=0.33\hsize]{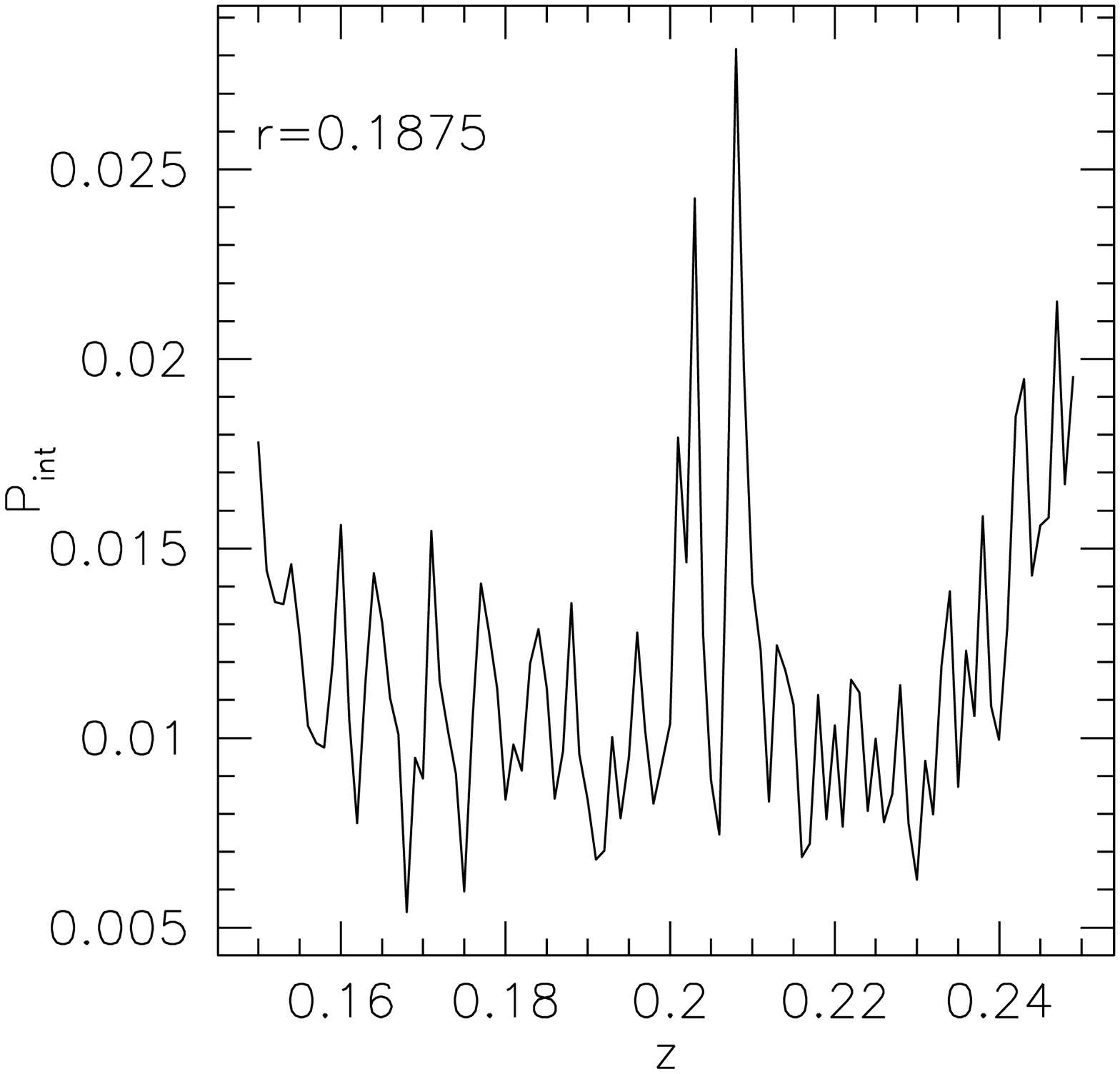}
\caption{Total integrated power $P_\mathrm{int}$ as function of
  redshift for five different distances $r$ (in units of $h^{-1}$Mpc)
  from the centre of the main cluster clump, as indicated in each
  panel.}
\label{figure:mergpspec}
\end{figure*}

\subsection{Dynamical state of the cluster}

\begin{figure*}
\centering
  \includegraphics[width=1.0\hsize]{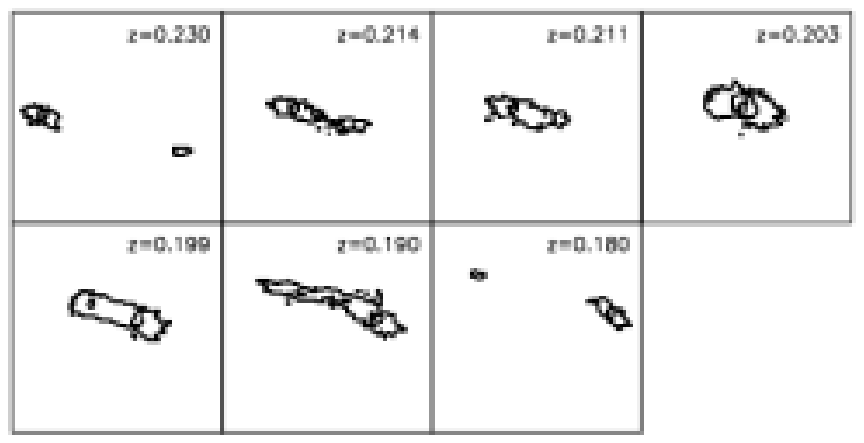}
\caption{Critical lines of the numerically simulated galaxy cluster
  (``optimal'' projection) at the same redshifts shown in
  Fig.~\ref{figure:mergsurfmaps}. The scale of each panel is $375''$.}
\label{figure:numclines}
\end{figure*}

We show in Fig.~\ref{figure:mergsurfmaps} some projected density maps
of the simulated cluster at several redshifts during the merger. As is
easily seen, a large substructure crosses the cluster centre at $z
\sim 0.2$. The (comoving) side length of the displayed region is
$3\,h^{-1}\,\mathrm{Mpc}$. We quantify the dynamical state of the
cluster using the multipole expansion technique of the surface density
field discussed by \citet{ME03.1}. Briefly, we place the origin of our
reference frame on the centre of the main cluster clump. Starting from
the particle positions in the numerical simulations, we compute the
surface density $\Sigma$ at discrete radii $r_n$ and position angles
$\phi_k$ taken from $[0,1.5]\,h^{-1}\,\mbox{Mpc}$ and $[0,2\pi]$,
respectively. For any $r_n$, each discrete sample of data
$\Sigma(r_n,\phi_k)$ is expanded into a Fourier series in the position
angle,
\begin{equation}
  \Sigma(r_n,\phi_k)=\sum_{l=0}^\infty\,S_l(r_n)
  \mathrm{e}^{-\mathrm{i}l\phi_k}\;,
\label{eq:expansion}
\end{equation}
where the coefficients $S_l(r_n)$ are given by
\begin{equation}
  S_l(r_n)=\sum_{k=0}^\infty\,\Sigma(r_n,\phi_k)
  \mathrm{e}^{\mathrm{i}l\phi_k}\;,
\end{equation}
and can be computed using fast-Fourier techniques. 

We define the power spectrum $P_n(l)$ of the multipole expansion $l$
as $P_n(l)=|S_l(r_n)|^2$. As discussed by \citet{ME03.1}, the amount
of substructure and the degree of asymmetry in the mass distributions
of the numerically simulated cluster at any distance $r_n$ from the
main clump can be quantified by defining an integrated power
$P_\mathrm{int}(r_n)$ as the sum of the power spectral densities over
all multipoles, from which we subtract the monopole and the quadrupole
in order to remove the axially symmetric and elliptical contributions,
\begin{equation}
  P_\mathrm{int}(r_n)=\sum_{l=0}^\infty\,P_n(l) - P_n(0) - P_n(2)\;.
\end{equation}
This quantity measures the deviation from an elliptical distribution
of the surface mass density at a given distance $r_n$ from the cluster
centre.

The results are shown in Fig.~\ref{figure:mergpspec}: the total
integrated power is shown as a function of redshift for five different
distances $r$ from the centre of the main cluster body. The peaks in
each plot arise when the infalling substructure enters or exits a
circle of radius $r$. Therefore, through the location of the peaks, we
can determine with precision the distance between the two merging
clumps at each redshift. The merger occurs between
$z_\mathrm{in}=0.250$ and $z_\mathrm{fin}=0.150$. In order to follow
in detail this event, we pick $N_\mathrm{snap}=100$ simulation
snapshots equidistant in redshift between $z_\mathrm{in}$ and
$z_\mathrm{fin}$. The redshift interval between two consecutive
snapshots therefore is $\Delta z=0.001$. In a $\Lambda$CDM model, this
corresponds to a time interval of approximately $\Delta t \sim 10$
Myr.

At redshift $z \sim 0.3$, the virial mass of our numerical halo is
$\sim 7 \times 10^{14} h^{-1} M_{\odot}$. In order to increase the
lensing efficiency and thus to reduce uncertainties in the numerically
determined arc cross sections, we artificially rescale the cluster
mass by multiplying the particle masses by a factor $f=2.5$. Recalling
that the virial radius $R_\mathrm{vir}$ of a halo scales as
$R_\mathrm{vir} \propto M^{1/3}$, where $M$ is the virial mass, to
obtain a halo of mass $f \times M$ which is dynamically stable, we
also need to rescale the distances by a factor $f^{1/3}$. This means
that while the three-dimensional density $\rho$ remains fixed, the
halo surface density is enhanced by a factor $f^{1/3}$. For this
reason we expect that, increasing its mass, the numerical cluster will
become much more able to produce strong lensing effects.

\subsection{Lensing simulations}
\label{section:lenssim}

\begin{figure*}
\flushleft
  \includegraphics[width=0.24\hsize]{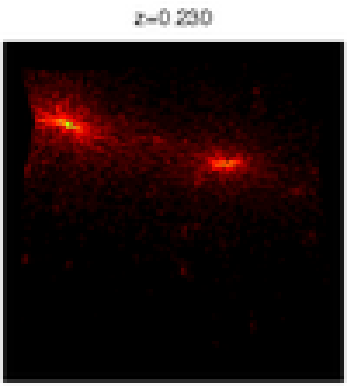}
  \includegraphics[width=0.24\hsize]{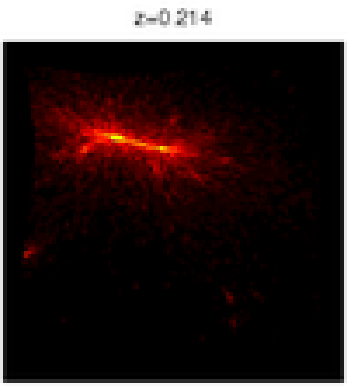}
  \includegraphics[width=0.24\hsize]{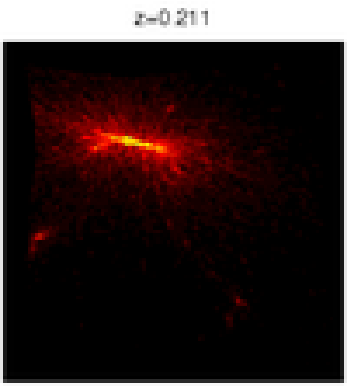}
  \includegraphics[width=0.24\hsize]{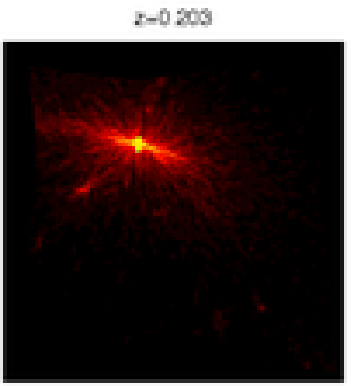}
  \includegraphics[width=0.24\hsize]{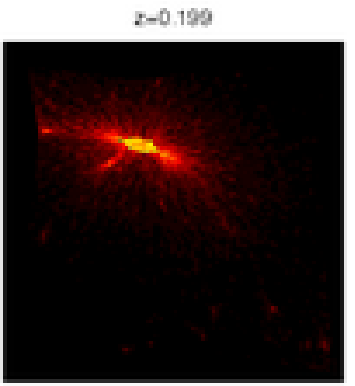}
  \includegraphics[width=0.24\hsize]{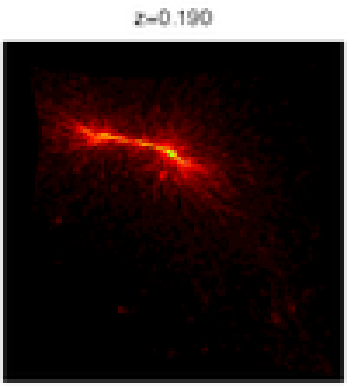}
  \includegraphics[width=0.24\hsize]{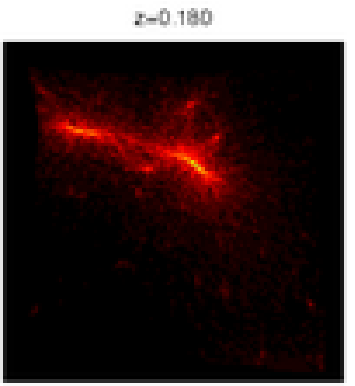}

\centering
  \includegraphics[width=0.30\hsize]{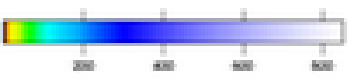}
\caption{Maps of the magnification on the source plane produced by the
  numerically simulated galaxy cluster at the same redshifts shown in
  Fig.~\ref{figure:mergsurfmaps}. The scale of each panel is $1.5 \
  h^{-1}$Mpc.}
\label{figure:mergmupdsmaps}
\end{figure*}

Using each of the $N_\mathrm{snap}$ snapshots, we perform ray-tracing
simulations. Our method is described in detail elsewhere
\citep{ME00.1,ME01.1,ME03.2,ME03.1}, but some parameters differ from
the simulations described earlier.

For each snapshot, we select again from the simulation box a cube of
$3 \ h^{-1}$Mpc side length, containing the high-density region of the
cluster. The centre of the cube is defined such that the selected
region contains both merging clumps. A third smaller substructure
enters the field at $z\sim 0.23$. In the lensing simulations performed
using the snapshots at higher redshift the effect of this small clump
of matter is not taken into account, because it is out of the
considered region. We checked its influence on the cluster strong
lensing properties by selecting a larger region. We found that the
effect of this substructure is negligible when its distance from the
main clump is larger than $\sim 1.5 h^{-1}$Mpc.

We then determine the three-dimensional density field $\rho$ of the
cluster from the particle positions, by interpolating the mass density
within a regular grid of $256^3$ cells, using the \emph{Triangular
Shaped Cloud} method \citep{HO88.1}. We then produce surface density
fields $\Sigma$ by projecting $\rho$ along three coordinate axes. We
chose our reference frame such that one axis is perpendicular to the
direction of merging. By projecting along this coordinate axis, we
minimize the impact parameter of the infalling substructure with
respect to the main cluster clump.  Hereafter this projection will be
called ``optimal''.  For comparison, we also investigate the
projection along a second axis, where the minimal distance between the
merging clumps is never smaller than $\sim 250 \ h^{-1}$kpc. The third
projection, which is not interesting for the purpose of this paper,
since the substructure moves almost exactly along the line of sight,
will be neglected in the following analysis.

The lensing simulations are performed by tracing a bundle of $2048
\times 2048$ light rays through the central quarter of the lens plane
and computing for each of them the deflection angle. Then, a large
number of sources is distributed on the source plane. We place this
plane at redshift $z_\mathrm{s}=1$. Keeping all sources at the same
redshift is an approximation justified for the purposes of the present
case study, but the recent detections of arcs in high-redshift
clusters \citep{ZA03.1,GL03.1} indicate that more detailed simulations
will have to account for a wide source redshift distribution. 

\begin{figure*}
\centering
  \includegraphics[width=1.0\hsize]{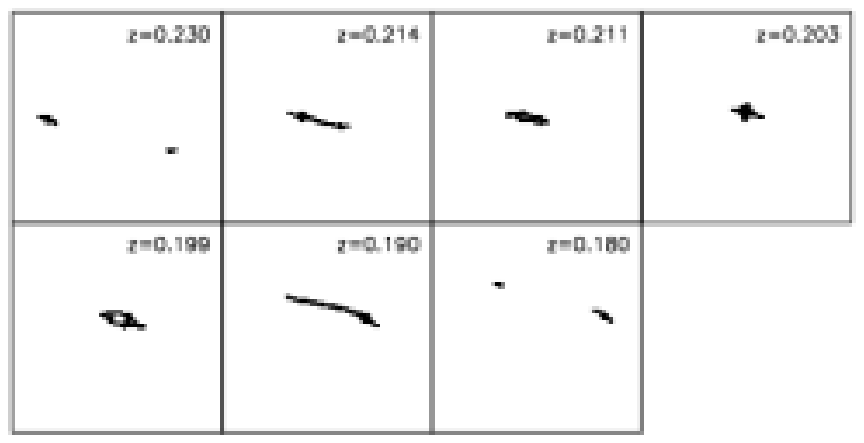}
\caption{Caustics of the numerically simulated galaxy cluster at the
  same redshifts as shown in Fig.~\ref{figure:mergsurfmaps}. The scale
  of each panel is $\sim 375''$. Note the appearance of thin, long
  caustic structures which are not seen in the analytic model shown in
  Fig.~\ref{figure:caustl} due to the lack of internal asymmetries and
  external shear.}
\label{figure:numcaustlines}
\end{figure*}

The sources are elliptical with axis ratios randomly drawn from
$[0.5,1]$. Their equivalent diameter (the diameter of the circle
enclosing the same area of the source) is $r_\mathrm{e}=1''$. Unlike
previous studies, we do not distribute the elliptical sources on the
central quarter of the source plane, since we need to investigate a
region large enough to contain the caustics of both the merging
clumps. Nevertheless, we keep the same spatial resolution of the
source grids in the previous simulations, because sources are
initially distributed on a regular grid of $64 \times 64$ instead of
$32 \times 32$ cells. Then, we adaptively increase the source number
density in the high magnification regions of the source plane by
adding sources on sub-grids whose resolution is increased towards the
lens caustics. This increases the probability of producing long arcs
and thus the numerical efficiency of the method. In order to
compensate for this artificial source-density increase, we assign for
the following statistical analysis to each image a statistical weight
proportional to the area of the grid cell on which the source was
placed. We refine the source grid four times, and the total number of
sources is typically $\sim30,000$.

Using the ray-tracing technique, we reconstruct the images of
background galaxies and measure their length and width. Our technique
for image detection and classification was described in detail by
\citet{BA94.1} and adopted by \citet{BA98.2} and
\citet{ME00.1,ME01.1,ME03.2,ME03.1}. It results in a catalogue of
simulated images which is subsequently analysed statistically.

\section{Results}

In the following sections we show how the lensing properties of our
numerical model change during the merger. We focus on the ``optimal''
projection of the numerical cluster, i.e. the projection along the
axis perpendicular to the direction of the merger, where the effects
of the merger are expected to be strongest. We refer to the second
projection only in passing.

\subsection{Critical lines and caustics}

As expected, the critical lines and the caustics of the numerical lens
evolve strongly during the merger.

First, we discuss the results for the ``optimal'' projection.  We show
the critical lines at some relevant redshifts in
Fig.~\ref{figure:numclines}. At redshift $z=0.230$, the main cluster
clump and the infalling substructure develop separate critical
lines. The largest mass concentration also produces a small radial
critical line (enclosed by the more extended tangential critical
line). As the merger proceeds, the tangential critical lines are
stretched towards each other. As discussed in
Sect.~\ref{section:mergexpectations}, this is due to the increasing
shear in the region between the mass concentrations. The critical
lines merge approximately at redshift $z=0.214$. After that, there
exists a single critical line, which, after a short phase of
shrinking, expands isotropically while the two clumps overlap. This
happens at $z\sim 0.203$.

Note that the radial critical line grows during the merging event,
reaching the maximum extension when the clumps are exactly aligned and
the surface density is highest. Indeed, in order to develop a radial
critical line, the lens must reach a sufficiently high central surface
density.

When the substructure moves to the opposite side of the main cluster
body, the tangential critical line stretches again and reaches its
maximum elongation at $z \sim 0.190$. Then, separate critical lines
appear around each clump. Their size decreases for decreasing $z$
because both the shear and the convergence between the two mass
concentration decrease as their distance grows. Similarly, the radial
critical lines shrink.

During the merger phase, the magnification pattern on the source plane
changes as shown in Fig.~\ref{figure:mergmupdsmaps}. The highest
magnifications are reached during the phase of maximum overlap, but
the extent of the highest magnification regions is largest at
redshifts $z=0.214$ and $z=0.190$. Therefore, at these redshifts even
the caustics are most elongated. In Fig.~\ref{figure:numcaustlines} we
show the caustics at the same redshifts as in
Fig.~\ref{figure:mergmupdsmaps}. Their evolution reflects that of the
critical lines. Before redshift $z=0.214$, two separated caustics
exist; as the distance between the two clumps shrinks, their
elongation in the direction of merging grows. Then the caustics merge
into a single line, shrink along their long axis and widen
perpendicular to it. When the two mass concentrations overlap, it has
a diamond shape with four pronounced cusps. Finally, when the distance
between the substructure and the main cluster clump grows again, the
caustic shrinks, elongates in the direction of relative motion of the
substructure, and finally splits into two small separate caustics.

\begin{figure*}
\centering
  \includegraphics[width=1.0\hsize]{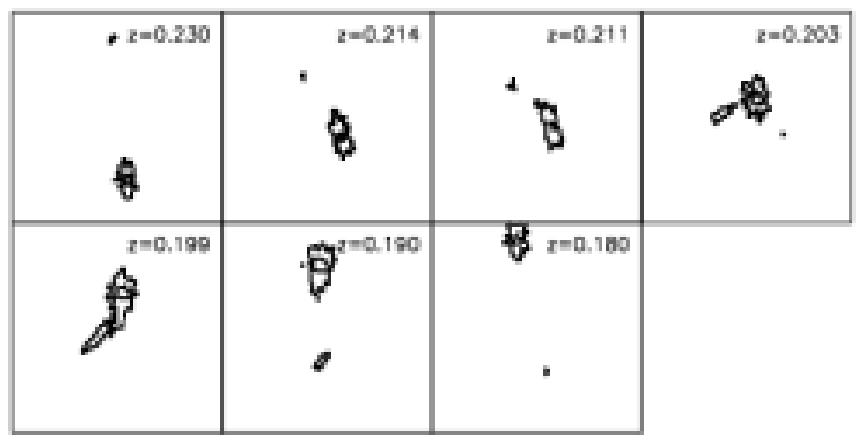}
\caption{Critical lines of the numerically simulated galaxy cluster
  (second projection) at the same redshifts shown in
  Fig.~\ref{figure:mergsurfmaps}. The scale of each panel is $375''$.}
\label{figure:critl2}
\end{figure*}

In the second projection we have considered, the substructure does not
pass through the cluster centre. In this case its distance from the
main cluster clump is always $\ga 250 h^{-1}$kpc. We show in
Fig.~\ref{figure:critl2} the critical lines for this projection. As in
the case of the ``optimal'' projection, while the merger proceeds, the
critical lines of the two merging clumps are stretched towards each
other by the effect of the increasing shear in the region between the
two mass concentrations. The largest elongation is reached at redshift
$\sim 0.198$. However, the substructure does not cross the region
enclosed by the critical lines of the main cluster clump and therefore
the critical lines do not shrink in the direction of merging and never
expand isotropically. After reaching the maximal elongation, it
shrinks while the distance between the two clumps grows. The caustics
(not shown) reflect the same evolution.

\subsection{Tangential and radial arcs}
\label{section:mergarcstat}

\begin{figure*}
\flushleft
  \includegraphics[width=0.49\hsize]{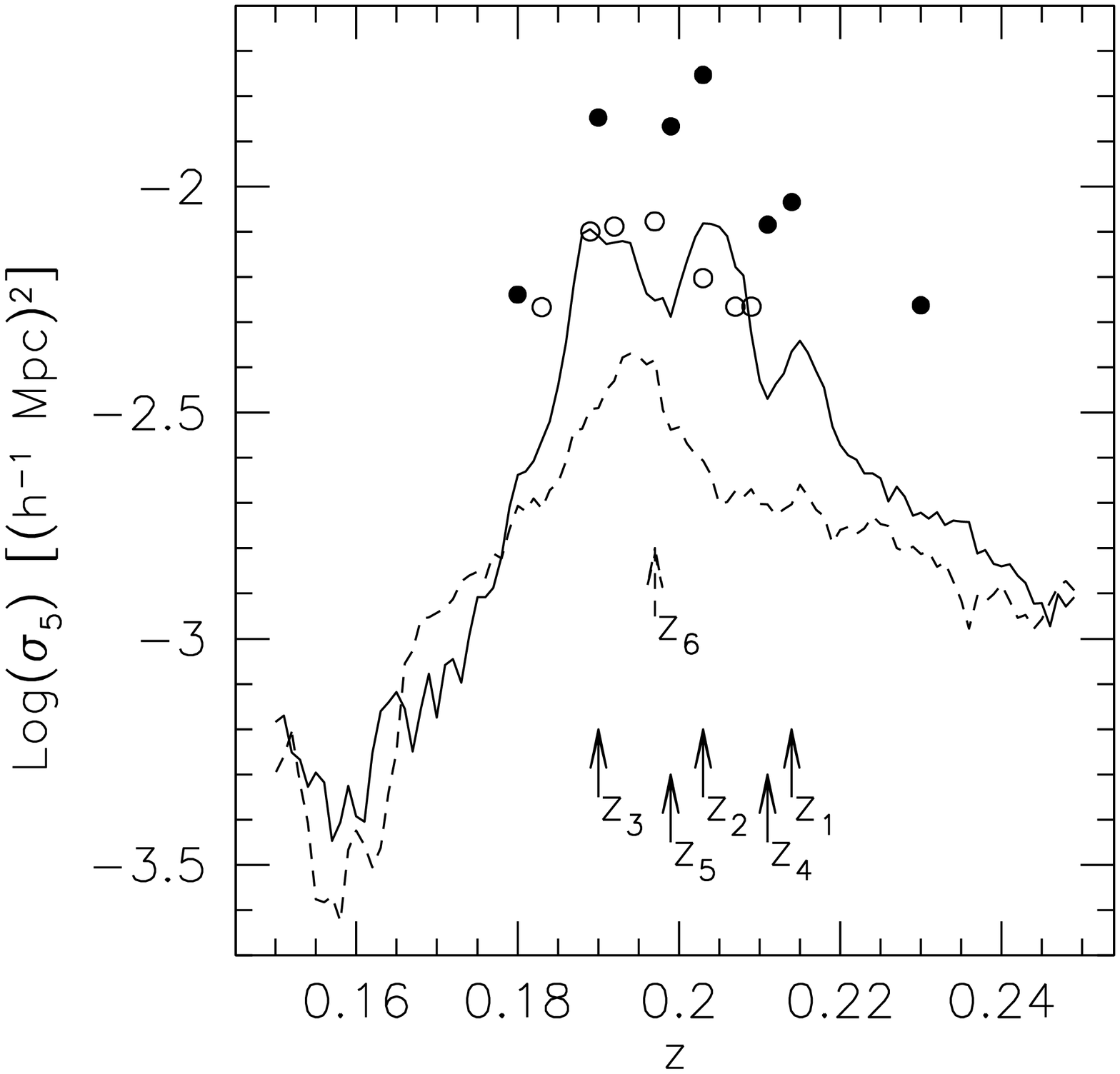}
  \includegraphics[width=0.49\hsize]{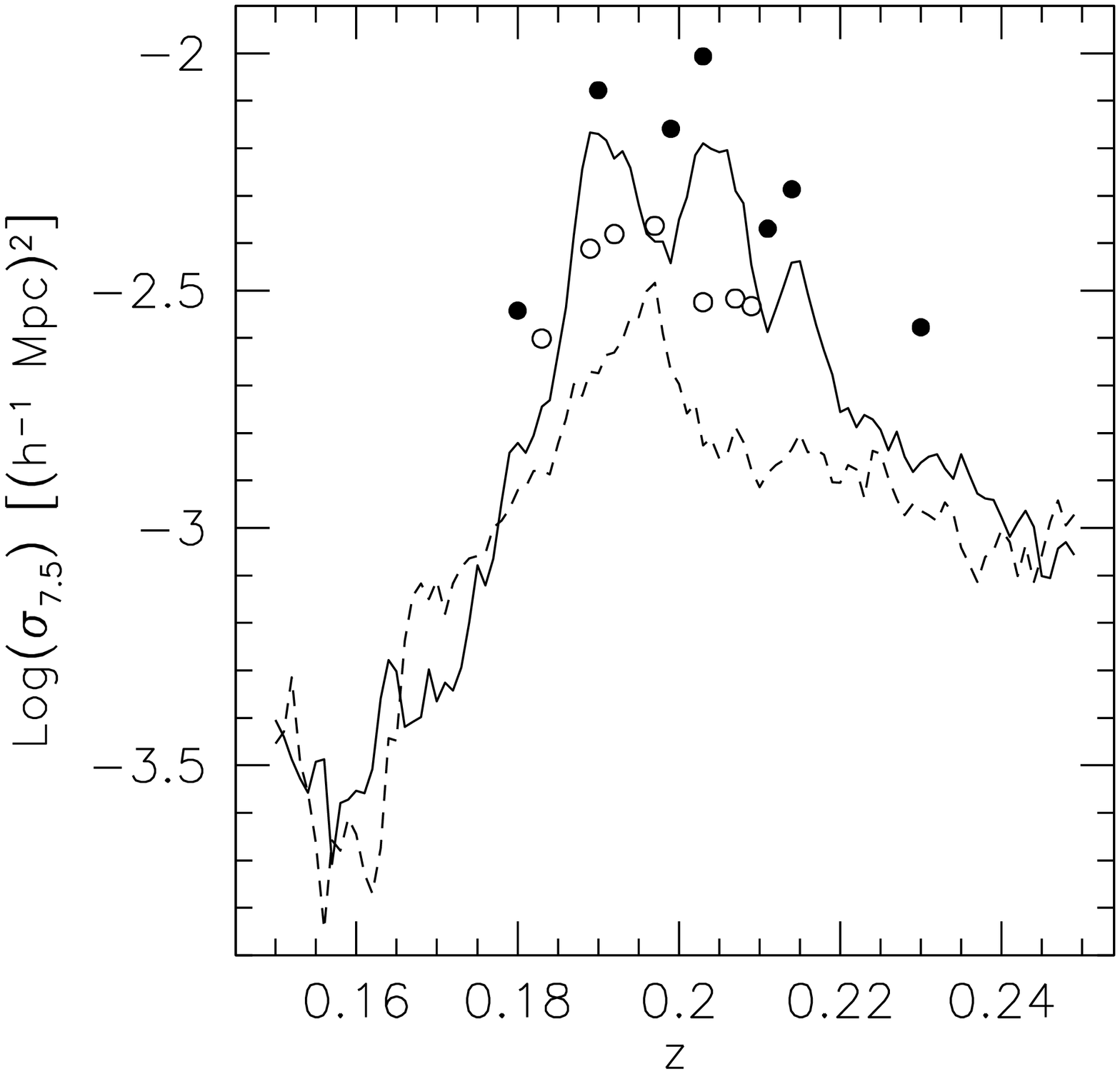}
  \includegraphics[width=0.49\hsize]{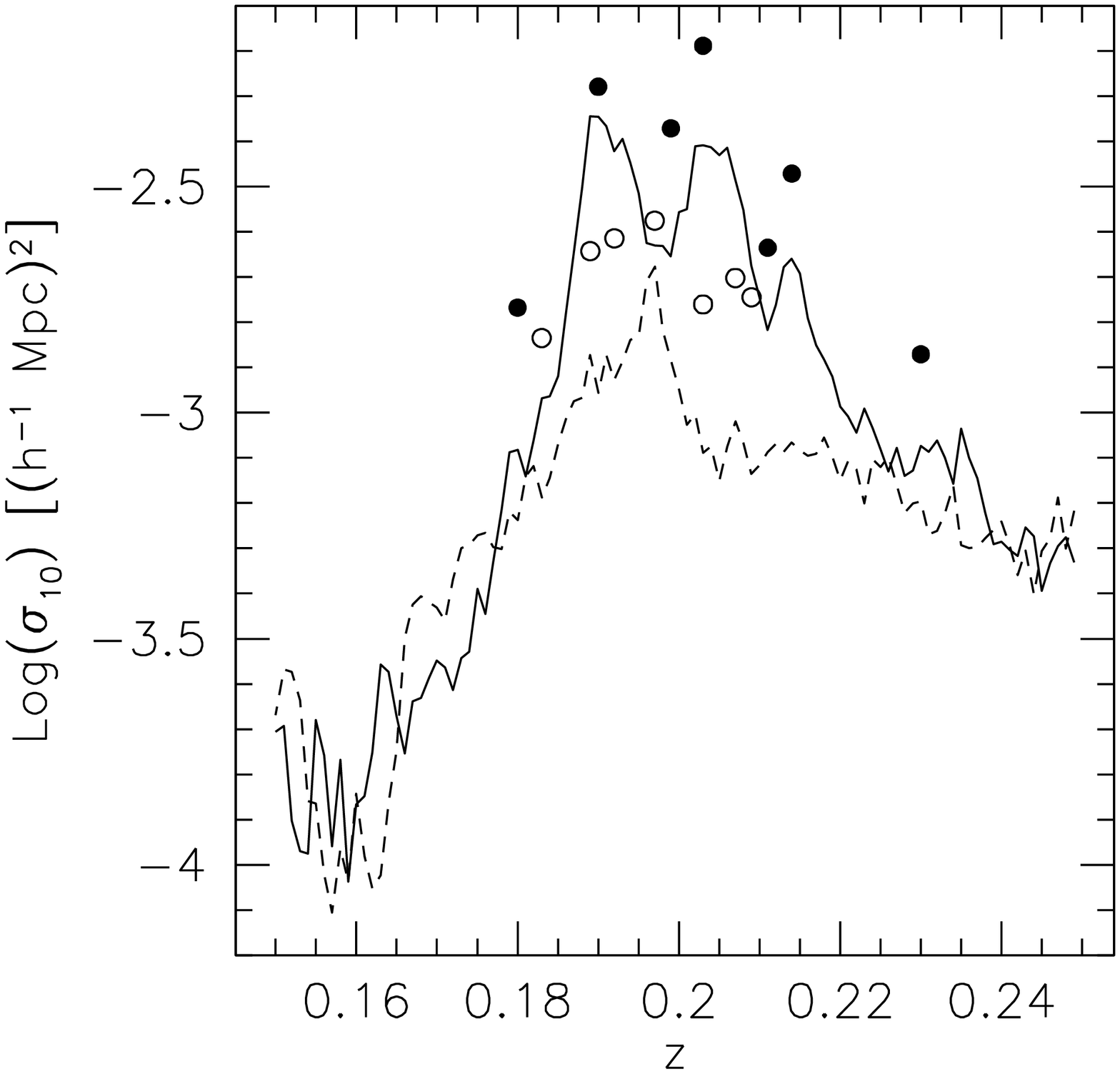}
  \includegraphics[width=0.49\hsize]{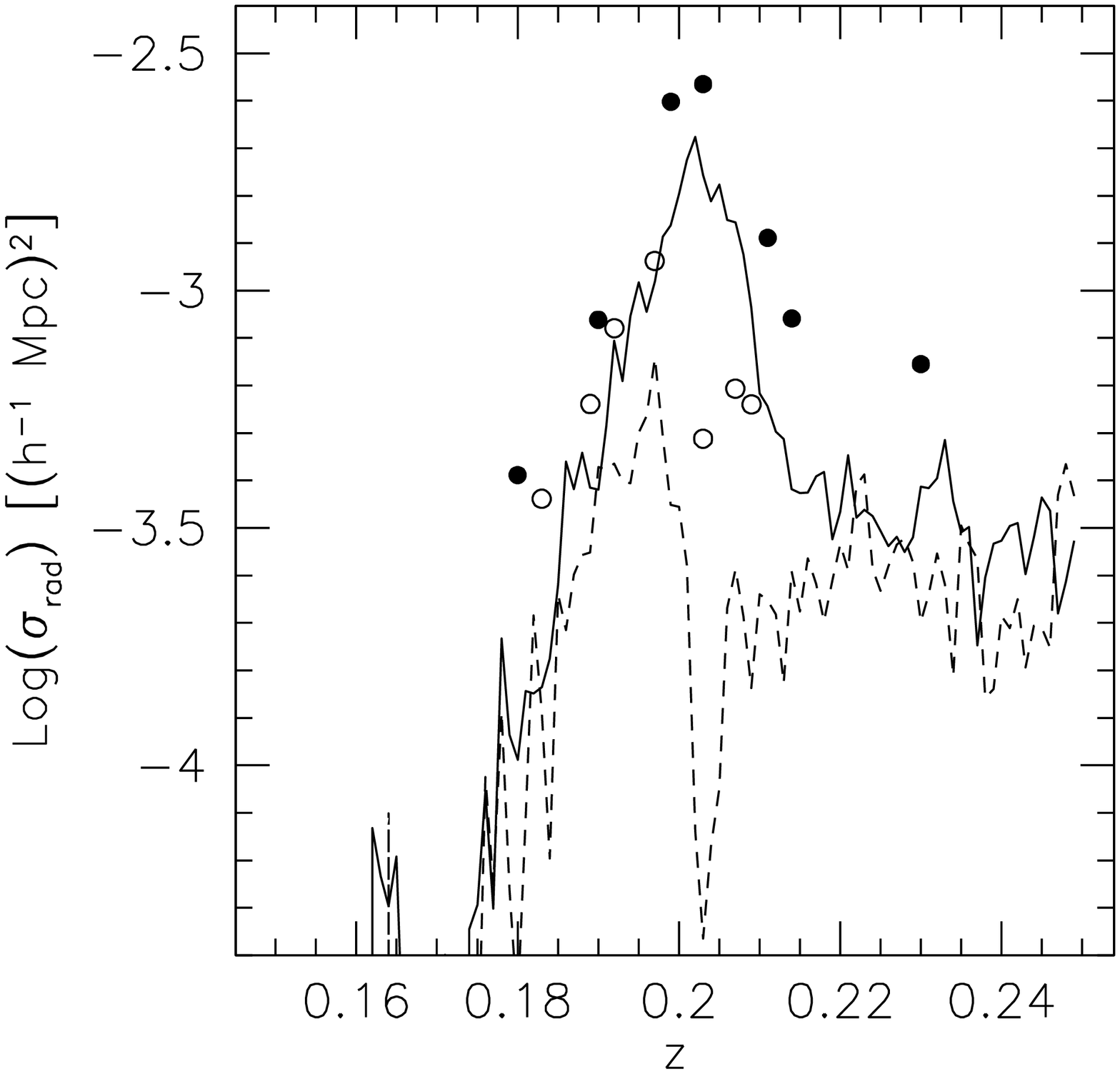}
\caption{Lensing cross sections for tangential and radial arcs as
  functions of redshift. Top left panel: arcs with length-to-with
  ratio $L/W>5$, $\sigma_5$; top right panel: arcs with length-to-with
  ratio $L/W>7.5$, $\sigma_{7.5}$; bottom left panel: arcs with
  length-to-with ratio $L/W>10$, $\sigma_{10}$; bottom right panel:
  radial arcs, $\sigma_\mathrm{rad}$. The solid lines refer to the
  ``optimal'' projection, i.e. where the substructure crosses the
  centre of the main cluster clump, while the dashed lines refer to
  the second projection of the cluster; in both cases the sources are placed 
  at redshift $z_\mathrm{s}=1$. Note the three peaks in the
  top and bottom left panels; their origin is discussed in the text.
  Cross sections computed considering sources placed at $z_\mathrm{s}=2$
  are shown for the ``optimal'' and second projections by filled and 
  open circles, respectively.
}
\label{figure:mergcsec}
\end{figure*}

Since the extent of the high-magnification regions and caustics
changes during the infall of the substructure onto the main cluster
clump as shown in the previous section, we expect that the lensing
efficiency for producing both tangential and radial arcs changes
accordingly. We quantify the influence of merging on the cluster's
strong lensing efficiency by measuring its cross sections for
different arc properties.

First, we consider tangential arcs, which can be identified among the
lensed images due to their large length-to-width ratio $L/W$. By
definition, the lensing cross section is the area on the source plane
where a source must be located in order to be imaged as an arc with
the specified property. As explained in Sect.~\ref{section:lenssim},
each source is taken to represent a fraction of the source plane. We
assign a statistical weight of unity to the sources which are placed
on the sub-grid with the highest resolution. These cells have area
$A$. The lensing cross section is then measured by counting the
statistical weights of the sources whose images satisfy a specified
property. If a source has multiple images with the required
characteristics, its statistical weight is multiplied by the number of
such images. Thus, the formula for computing cross sections for arcs
with a property $p$, $\sigma_\mathrm{p}$, is
\begin{equation}
  \sigma_\mathrm{p}=A\,\sum_i\,w_i\,n_i\;,
\end{equation}
where $n_i$ is the number of images of the $i$-th source satisfying
the required conditions, and $w_i$ is the statistical weight of the
source. Using this method, we compute the lensing cross sections for
arcs with $L/W \ge 5, 7.5$ and $10$, respectively.

The results are shown in the first three panels of
Fig.~\ref{figure:mergcsec}. The solid lines refer to the ``optimal''
projection. All curves exhibit the same redshift evolution. Their main
properties can be summarized as follows:
\begin{itemize}
\item The cross sections grow by a factor of two between $z\sim 0.240$
  and $z \sim 0.220$, as shown in Fig.~\ref{figure:mergpspec}. Then,
  the cross section further increases by a factor of five between
  $z\sim 0.220$ and $z\sim 0.200$, i.e. within $\sim 0.2$ Gyr.
    
\item The curves have three peaks, located at redshifts $z_1=0.214$,
  $z_2=0.203$ and $z_3=0.190$. The peaks at $z_1$ and $z_3$ correspond
  to the maximum extent of caustics and critical curves along the
  merging direction \emph{before} and \emph{after} the moment when the
  merging clumps overlap; the peak at $z_2$ occurs when the distance
  of the infalling substructure from the merging clump is minimal.

\item Two local minima arise between the three maxima at redshifts
  $z_4=0.211$ and $z_5=0.199$, where the cross sections are a factor
  of two smaller than at the peaks. At these redshifts, the critical
  lines have shrunk along the merging direction and the caustics are
  less cuspy than at $z_2$.

\item The cross sections reduce by more than one order of magnitude
  after $z=0.190$. At redshift $z\sim 0.180$, when the distance
  between the merging clumps is $\ga 1.5 \ h^{-1}$Mpc, the cross
  section sizes are comparable to those before $z \sim 0.240$.
\end{itemize}

\begin{figure*}
\centering
{\bf \large ``Optimal'' projection} \\
 \includegraphics[width=0.70\hsize]{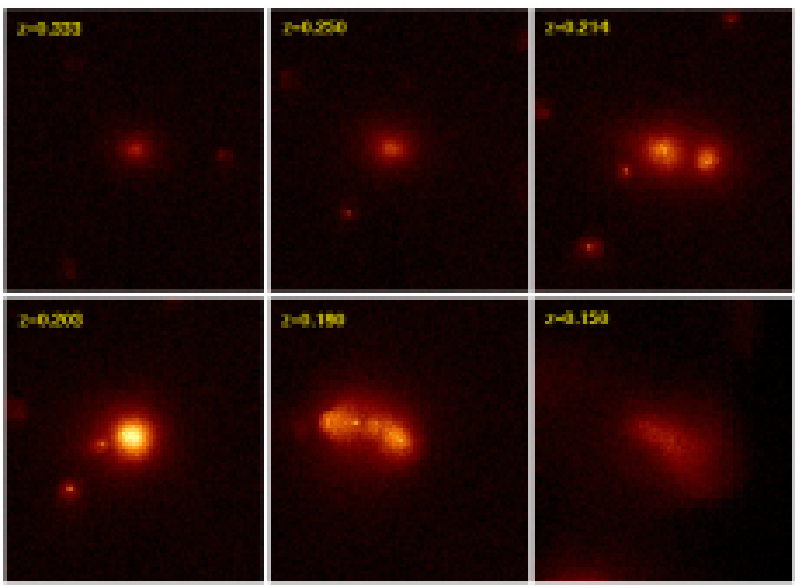}\\
{\bf \large Second projection} \\
 \includegraphics[width=0.70\hsize]{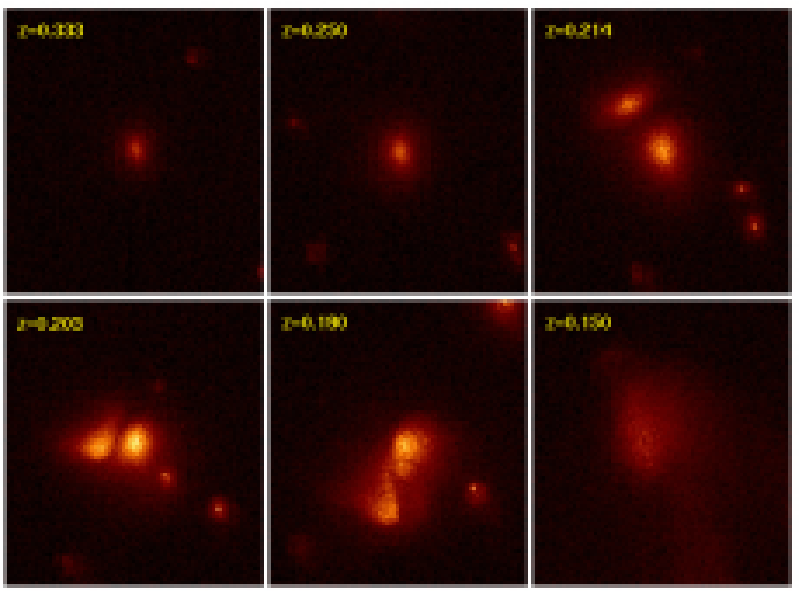}
\caption{Simulated observations with the \emph{Chandra} satellite of
  the X-ray emission by our numerically modelled galaxy cluster. The
  plots show the distributions of detection events on the CCD
  ACIS-S3. The exposure time is $30,000$ sec, while the scale of each
  figure is $\sim 2.1 \ h^{-1}$Mpc. The upper and lower panels refer
  to the ``optimal'' and to the second projections, respectively.}
\label{figure:xsimul}
\end{figure*}

Therefore, during the merger, our simulated cluster becomes extremely
more efficient in producing tangential arcs. The infalling
substructure starts affecting the cross sections for long and thin
arcs when its distance from the main cluster clump is approximately
equivalent to the cluster virial radius ($\sim 1.5 \ h^{-1}$Mpc), and
the largest effects are seen at three different times:
\begin{itemize}
\item When the critical lines (and the corresponding caustics) merge,
  i.e. when the shear between the mass concentrations is sufficient to
  produce the largest elongation of the critical lines along the
  direction of merging. This happens $\sim 100$ Myr before and after
  the substructure crosses the cluster centre;
\item when the two clumps overlap, i.e. when the projected surface
  density or convergence is maximal, producing the largest {\em
  isotropic} expansion of the critical lines and of the caustics.
\end{itemize}

A different behaviour is found in the second projection. The results
are shown again in the first three panels of
Fig.~\ref{figure:mergcsec} as dashed lines. In this case the critical
lines and caustics reach a maximal elongation at $z_6=0.198$ and
shrink at lower redshifts. Again the lensing cross sections reflect
the evolution of the critical curves and caustics: they reach a
maximum extent at $z_6$ and then their size decreases. A very
important result is that, even if the merging clumps never get closer
than $\sim 250 \ h^{-1}$kpc, the cross sections still grow by roughly
a factor of five within approximately $50$ Myr.
 
We now consider the effects of merging on the cross section for radial
arcs. These are identified from the complete sample of distorted
images using the technique described in \citet{ME01.1}. It consists in
selecting those arcs for which the measured radial magnification at
their position exceeds a given threshold.

The cross section for this type of arcs as a function of redshift is
shown in the fourth panel of Fig.~\ref{figure:mergcsec}. In both
projections, at redshifts $z \ga 0.22$, the cross section for radial
arcs keeps constant and fluctuates around $10^{-3.5} \
h^{-2}$Mpc$^2$. Then, in the ``optimal'' projection, it grows by a
factor of five, reaching the highest value at $z=0.203$. This is the
same redshift where the cross sections for tangential arcs peak. The
enhancement of the convergence due the overlapping of the merging
clumps thus makes the cluster substantially more efficient for
producing radial arcs.

Then, the cross section rapidly drops to zero for smaller
redshifts. Note that the redshift interval where our cluster is very
efficient for producing radial arcs is quite small ($\Delta
z_\mathrm{rad}\la0.04$). Radial arcs have so far been reported in only
five galaxy clusters (MS~2137, \citealt{FO92.1}; A~370,
\citealt{SM96.1}; MS~0440, \citealt{GI98.1}; AC~114, \citealt{NA98.1};
A~383, \citealt{SM01.1}). Our results contribute to explaining why
radial arcs are so rare: if they form preferentially during the
crossing of large substructures through the cluster centre, when the
central lens surface density is higher, the visibility window of such
events is very narrow ($\la 100$ Myr per merger).

Note that, in the second projection (dashed line), the cross section
for radial arcs has some strong fluctuations during the merger but
never reaches very high values. As for the ``optimal'' projection, it
peaks at the same redshift where the cross section for tangential arcs
is largest, but it changes by less than a factor of two with respect
to redshifts $z \ga 0.21$. This confirms that only large enhancements
of the central surface density produce a substantial increase of the
cluster's efficiency for forming radial arcs.

As mentioned before, assuming a single source redshift is justified
for the purpose of this work. Indeed, the critical surface density
changes by $\sim 10$ per cent when moving the sources from redshift
one to redshift two, while keeping the lens redshift at $z_{\rm l}
\sim 0.2$. Thus the extent of the critical lines and caustics (and
consequently of the lensing cross sections) is not expected to change
very much for sources at redshifts above unity. Nevertheless, we show
in Fig.~\ref{figure:mergcsec} the cross sections obtained by placing
the source plane at $z_{\rm s}=2$ for some characteristic snapshots of
our simulations. Filled and open circles refer to the ``optimal'' and
the second projections, respectively. As expected the relative change
in the amplitude of the cross sections is modest (less than a factor
of $2$) for both tangential and radial arcs. Moreover, the amplitude
of the fluctuations induced by the merger event on the lensing cross
sections seems to depend very weakly on the source redshift.

\section{Observational implications}

Our results show that cluster mergers could play an important role for
arc statistics. In particular, since the lensing efficiency grows by
one order of magnitude during mergers, they may offer a solution for
the \emph{arc statistics problem}.

It is quite important to notice that mergers might have some other
important observational implications to account for.  In fact the
largest sample of clusters used for arc statistics studies
\citep{LU99.1} was selected in the X-ray band, where the luminosity is
due to bremsstrahlung emission. This is very sensitive to the
dynamical processes going on in the cluster, since it is proportional
to the square gas density. Therefore, we expect that the cluster X-ray
luminosity has large variations during a merging phase.

In Fig.~\ref{figure:xsimul} we show simulated X-ray observations of
our cluster. These were obtained by using X-MAS, a code for simulating
data taken with the \emph{Chandra} satellite \citep{GA03.1}. The
cluster emissivity is calculated integrating over the gas density and
temperature distribution within the cluster simulation, adopting the
MEKAL plasma model \citep{KA93.2} with a metallicity of
$0.3\,Z_\odot$. The upper panels present the results in the
``optimal'' projection at six different redshifts. The upper left
panel shows the cluster at $z=0.333$, before the merger starts. The
second panel shows the cluster at $z=0.250$, when the virial regions
merge. The third, fourth and fifth panels show the cluster just
\emph{before}, \emph{at} and just \emph{after} the maximum overlap of
the merging clumps, respectively. Finally, in the sixth panel, the
cluster is observed after the end of the merger. Since the colour
scale is the same for all images, it can be easily seen that the X-ray
luminosity is highest when the merging clumps are closest. Simular
results for the second projection are shown in the bottom panels.

\begin{figure}
\centering
  \includegraphics[width=0.7\hsize]{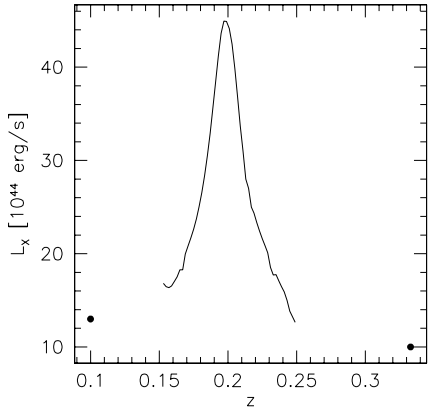}
\caption{X-ray luminosity $L_x$ of the numerically simulated cluster
  as function of redshift. The curve is accurately sampled in the
  redshift range $z=0.15 \div 0.25$. The X-ray luminosity has been
  measured even at $z=0.1$ and $z=0.333$, where is indicated by the
  filled circles.}
\label{figure:xlumz}
\end{figure}

In Fig.~\ref{figure:xlumz} we show the observed X-ray luminosity of
the numerical cluster as function of redshift. The curve has a narrow
and almost symmetric peak located at $z\sim 0.200$. The X-ray
luminosity grows by more than a factor of four between $z\sim 0.300$
and $z\sim 0.200$, by roughly a factor of $\sim 2.5$ between $z\sim
0.230$ and $z\sim 0.200$ and by roughly a factor of $\sim 1.55$
between $z\sim 0.210$ and $z\sim 0.200$. The width at half maximum of
the peak is approximately $\Delta z \la 0.05$. Note that the X-ray
luminosity peak is wider than the maxima in the arc cross sections,
thus the X-ray emission increases earlier and decreases later than the
arc cross sections during the merger.

If a cluster sample is built by collecting all the objects with X-ray
luminosity $L_X$ larger than a given threshold, we expect that many
merging clusters are present among them, since they are stronger X-ray
emitters.  Since these are all very efficient clusters for producing
gravitational arcs, this could introduce a bias in the observationally
determined frequency of \emph{giant arcs}, which could become larger
than predicted by previous numerical lensing simulations in
$\Lambda$CDM model.  However, it is quite difficult to make more
robust conclusions here since only one cluster has been
analysed. Further investigations are needed here. In any case, our
results show that much caution must be applied when selecting clusters
for arc statistics studies through their X-ray emission (cf.~also
\citealt{BA96.2}).

\section{Summary and conclusions}

In this paper we have investigated how the lensing properties of a
galaxy cluster change during merging events. Similar dynamical
processes were not resolved in the previous lensing simulations but
they might play a relevant role for determining the strong lensing
efficiency of cluster lenses.

We address this problem first by using analytic models. When
simulating a collision between spherical haloes with NFW density
profiles, we find that both the critical lines and the caustics of the
lens system strongly evolve during the merger. This behaviour is
explained by the change of the shear and of the convergence induced by
the infalling clump. Indeed, while the distance between the merging
mass concentrations decreases, the shear intensity grows in the region
between the halo centres. The individual critical lines and caustics
of the main cluster clump and the infalling substructure are stretched
towards each other until they merge.

To obtain a quantitative estimate of the impact of mergers on the
lensing cross sections, we have studied the strong lensing properties
of a numerically simulated galaxy cluster. Within this numerical
model, a massive substructure falls onto the main cluster clump
between $z_\mathrm{in}=0.250$ and $z_\mathrm{fin}=0.150$. We have
studied the merger in detail, picking a large number of simulation
snapshots within this redshift range. The time separation between
consecutive snapshots is approximately $\sim 0.01$ Gyr. Two different
projections of the cluster where analysed: in the ``optimal''
projection, the substructure passes through the centre of the main
cluster clump, while in the second projection the distance between the
two mass concentrations is always larger than $\sim 250 \ h^{-1}$kpc.

The main results of this study can be summarized as follows:
\begin{itemize}
\item As expected from the results of the analytic tests, the shapes
  of critical lines and caustics substantially change during the
  merger. At the beginning, the two clumps develop individual critical
  lines and caustics. These are stretched towards each other while the
  distance between the mass concentrations decreases, because the
  intensity of the shear field grows in the region between the
  approaching clumps. In the ``optimal'' projection, when they merge,
  the resulting single critical line and caustic shrink along the
  merging direction and then expand isotropically, because of the
  increasing convergence. The same behaviour is observed when the
  substructure moves far away from the main cluster clump after
  crossing its centre. In the other projection the maximum extent of
  the critical lines is reached when the distance between the mass
  concentrations is such that the effect of the shear is
  largest. After that, the size of the critical lines drops.
\item In the ``optimal'' projection, the lensing cross sections for
  tangential arcs change by one order of magnitude during the
  merger. The effect of the infalling substructure starts to be
  relevant when its distance from the main cluster clump is $\sim 1.5\
  h^{-1}$Mpc.  The cross sections have three peaks located at the
  redshifts where the critical lines have the largest extent along the
  merging direction, or when the shear effect induced by the infalling
  substructure is largest, and at the redshift where the two clumps
  overlap and consequently the maximum convergence is
  reached. Although the effects of the merger on the lensing cross
  sections are important within a time interval of $\sim 1$ Gyr, the
  strongest impact is thus observed during the central part of the
  merging phase, on a time scale of $\sim 200$ Myr. In the second
  projection, the lensing cross section for long and thin arcs change
  by a factor of five within a time interval of $\sim 100$ Myr.
\item The numerical cluster is highly efficient in producing radial
  arcs only during the merger. The cross section for this type of arcs
  has only one peak, located at the redshift were the infalling
  substructure crosses the centre of the main cluster clump.
\end{itemize}
   
Thus, our results show that mergers have a strong impact on the strong
lensing efficiency of galaxy clusters. Since the lensing cross
sections for long and thin arcs change by one order of magnitude
during the mergers, these dynamical processes could be a possible
solution to the arc statistics problem.

This picture is in principle supported by the fact that samples of
clusters used in arc statistics studies are selected through their
X-ray luminosity, which is very sensitive to the dynamical processes
arising in the cluster. In particular, we expect that many merging
clusters are present in these samples, since they are strong X-ray
emitters. For example, \citet{RA02.1} estimate that the number of clusters
with luminosities $L_X>5 \times 10^{44} h^{-2}$ erg/sec can be
increased by a factor of 8.9 due to merger boosts.

In addition, by surveying clusters in the LCDCS and in the RCS,
\citet{ZA03.1} and \citet{GL03.1} have recently found a high incidence
of giant arcs in clusters at high redshift. Their results are
particularly interesting since a large number of clusters merging at
high redshift are predicted by the commonly accepted theory of
structure formation. In particular, \citet{GL03.1} speculate that a
subset of clusters with low mass and large arc cross sections may be
responsible for large numbers of arcs in distant clusters. One
possibility is that such ``super-lenses'' are clusters in the process
of merging.

Detailed conclusions are, however, pending on further studies
quantifying the frequency of mergers and the dependence on arc cross
sections on the detailed merger parameters, such as the impact
parameter of the collision, the mass ratio of the merging haloes and
so forth. Such studies are now under way.

In recent studies, \citet{WA03.1} and \citet{DA03.1} suggest that arc
statistics in $\Lambda$CDM models can be reconciled with the observed
abundance of gravitational arcs by adopting a broader distribution of
source redshifts. Certainly, cross sections can change substantially
for a cluster of a given mass depending on its dynamical state, which
makes earlier and current statements about the theoretical
expectations highly insecure. Moreover, numerous observational effects
need to be taken into account in addition for a reliable comparison
between numerical simulations and observations.

\section*{Acknowledgements}

This work has been partially supported by Italian MIUR (Grant 2001,
prot.~2001028932, ``Clusters and groups of galaxies: the interplay of
dark and baryonic matter''), CNR and ASI.  MM thanks EARA for
financial support and the Max-Planck-Institut f\"ur Astrophysik for
the hospitality during the visits when part of this work was done.  We
are grateful to A. Gardini for the possibility of using the code X-MAS
simulating Chandra observations, and to S. Matarrese, N. Dalal and
J. Kempner for useful discussions.

\bibliography{mergClusArc}
\bibliographystyle{mn2e}

\end{document}